\documentclass{aaEC}  

\usepackage{graphicx}

\usepackage{txfonts}

\usepackage[switch, modulo]{lineno}
\nolinenumbers
\makeatletter
\renewcommand*\aa@pageof{, page \thepage{} of 18}
\makeatother
\begin{document}

   \title{Testing gravity with wide binaries}

   \subtitle{3D velocities and distances of wide binaries from Gaia and HARPS}

   \author{R. Saglia \inst{1}
     \and
     L. Pasquini\inst{2,4}
          \and
          F. Patat \inst{2}
          \and 
          H.-G. Ludwig  \inst{3}
          \and
          R. Giribaldi \inst{4}
         \and 
         I. Leao \inst{5}
         \and
         J. R. de Medeiros \inst{5}
         \and
         Michael T. Murphy \inst{6}  
          \fnmsep\thanks{Based on Archival ESO data}
          }
         \institute{Max Planck Institute for Extraterrestrial Physics, Giessenbachstr. 85748 Garching, Germany\\
                       \email{saglia@mpe.mpg.de}
         \and
               ESO, 
               Karl Schwarzschild Strasse 2, D-85478 Garching bei M\"unchen 
             \and 
              Landessternwarte  Zentrum f\"ur Astronomie der Universit\"at Heidelberg, K\"onigstuhl 12, 69117 Heidelberg, Germany   
         \and 
	         INAF, Osservatorio Astrofisico di Arcetri, Largo E. Fermi, Firenze, Italy 
         \and
                  Departamento de Fisica, Universidade Federal do Rio Grande do Norte, 59078-970 Natal, RN, Brazil
        \and 
   Centre for Astrophysics and Supercomputing, Swinburne University of Technology, Hawthorn, Victoria 3122, Australia.
          }
    
   \date{Received, ; accepted, }

   \abstract
       {Wide Binaries (WBs) are interesting systems to test
    Newton-Einstein gravity in low potentials. The basic concept is to
    verify whether the difference in velocity between the WB
    components is compatible with what is expected from the Newton
    law.  }
   {Previous attempts, based solely on Gaia proper motion differences
     scaled to transverse velocity differences using mean parallax
     distances, do not provide conclusive results.  Here we add to the
     Gaia transverse velocities precise measurements of the third
     velocity component, the radial velocity (RV), in order to
     identify multiple stars, and to improve the reliability of the
     test by using velocity differences and positions in three
     dimensions.}
   {We mine the ESO archive for
   observations of WBs with the high precision HARPS spectrograph, and
   use these observations to search for RV variations,
   that would indicate the presence of additional stars in the
   system. We use the HARPS spectra to determine accurate RV
   difference between the WB components, correcting the
   observed velocities for gravitational redshift and convective
   shift. We exploit the Gaia distance distributions to determine the
   projected and intrinsic separations $s$ and $r$ and the
   3-dimensional velocity differences of the binaries.}
     {We retrieve 44 pairs observed with HARPS, most of them with
       numerous observations, spanning time baselines from week to
       years. A considerable fraction (27$\%$) of these pairs show
       sign of multiplicity or are not suitable for the test, and 32
       bona-fide WBs survive our selection. Their projected
       separation {\it s} is up to 14 kAU, or 0.06 parsec. The median
       renormalized unit weight error (RUWE)
       parameter for the final sample is of 0.975 (highest value 1.24)
       and the median RV variability is 10 m~s$^{-1}$ with standard deviation
       6 m~s$^{-1}$.  Gaia RVs are on average smaller by only 68 m~s$^{-1}$ from
       those determined with HARPS,
       with an RMS dispersion of 311 m~s$^{-1}$. We determine distances,
       eccentricities and position angles to reproduce the velocity
       differences according to Newton's law, finding reasonable
       solutions for all WBs but one, and with some
systems possibly too near pericenter and/or at too high inclination. }
     {We show that precise (and accurate) multiple RVs of WB
       candidates are a very powerful tool to make the WBs test of
       gravity more robust and reliable. These observations allow to
       minimize or eliminate one of the major limitations of previous
       tests, that is the presence of multiple systems, and make the
       comparison with theory straightforward, without the need to
       resort to complex simulations. Our (limited) number of WBs does
       not show obvious trends with separation or acceleration and is
       consistent with Newtonian dynamics. We
       are collecting a larger sample of this kind to robustly assess
       these results. }


   \maketitle

\section{Introduction}
\label{sec_Intro}

General Relativity (GR) and its Newtonian limit
stands as a cornerstone of physics, proven
across a vast range of accelerations. However, in regions with
extremely low accelerations, such as the outskirts of galaxies, the
inclusion of an additional component, Dark Matter (DM) is required.
While cold DM models effectively reproduce many astrophysical phenomena,
the quest for the elusive particle responsible for DM underscores the
importance of rigorously testing gravity against the predictions of
GR.  Wide Binaries (WB) have been proposed as an ideal laboratory for
such scrutiny, given that their gravitational accelerations
($\sim 10^{-10}$ m~s$^{-2}$) closely
resemble those found in the external regions of galaxies, see
\citet{Hernandez+2012}. The underlying test
is seemingly straightforward: assessing the degree to which WBs adhere
to Kepler's laws. Given the exceedingly long period of their orbits,
spanning many thousands of years, a complete reconstruction of the
orbits is impossible. A statistical approach is instead employed,
involving the measurement of velocity differences between the
components of WBs and subsequent comparisons with simulations. The
magnitude of this velocity difference, for a binary separation of 0.1
pc, or 20 thousand astronomical units (20 kAU), is 50--200 m~s$^{-1}$,
depending on eccentricity, which already sets the requirements on the
velocity precision of this experiment. This test has been recently
enabled by Gaia, thanks to its exquisite astrometry. Several groups
have compiled samples of nearby WBs and executed the test using solely
Gaia-derived data, focusing on 2D transverse velocities obtained from
parallaxes, proper motions and separations \citep{Hernandez2022,Hernandez+2023,
  Pittordis+2019, Pittordis2023, Chae+2023, Banik+2024, Chae+2024}. These works are
limited to use 2D velocities because, given the magnitude of the
expected velocities involved, Gaia's radial velocities (RVs) in general fall
short of the required precision for this particular
application. \citet{Chae2025} improves on the situation by selecting
the 312 WBs with the most precise available Gaia velocities, going
beyond the transverse analysis. The outcomes of these investigations
present a lack of consensus, with opposite conclusions drawn by
different groups. While \citet{Hernandez2022, Hernandez+2023, Chae+2023,
  Chae+2024, Chae2025} suggest evidence indicating a departure from GR for
separations exceeding approximately 3 kAU, \citet{Pittordis+2019,Pittordis2023,
  Banik+2024} argue that no deviations from GR are detected.  As a
reference, at 3 kAU separation, the velocity difference expected
according to Newton law for two solar mass stars in circular orbits is
(see below) $V = 770$ m~s$^{-1 }$ and the acceleration
$1.3\times10^{-9}$ m~s$^{-2}$. The ongoing discussion underscores the
complexity of the matter, which can be attributed to several factors,
all arising from the absence of precise RVs.  A primary complicating
factor is the presence of triple systems. When one of the two
components within the WBs is a binary system itself (with a separation
too small to be resolved by Gaia), the resulting velocity becomes
inflated \citep{Clarke2020}.
Thus, it is imperative to incorporate triple systems into
simulations involving the WB population.  Transitioning from 3D orbits
to 2D projections introduces an additional layer of
uncertainty. Furthermore, the potential lack of physical association
among some WBs in the sample could contribute to contamination. These
effects persist, despite various precautions taken by researchers in
sample creation to minimize them.  The divergence in conclusions among
the studies primarily stems from simulating the parent WB population
with different selections criteria,
magnifying the need for more accurate observational data.

As suggested by \citet{Pittordis2018}, we overcome these limitations by searching in the ESO archive for
HARPS \citep{Mayor+2003} observations of systems classified in
literature as wide binaries, finding 44 systems in the archive,
to obtain accurate measurements of their third (radial) velocity
dimension, to exclude the presence of triple systems and to make a full
3D velocity comparison. HARPS observations typically span long (up
to several years) time baseline, and thanks to the excellent RV precision (typically better than 10 m~s$^{-1}$) of this instrument, it is possible to determine
whether a hidden companion contaminates the system. Moreover, we
exploit the information coded by the Gaia parallaxes and their errors
to estimate the probability distributions of the WB distances and
intrinsic separations.

The paper is organized as follows. Section \ref{sec_method} explains
the method used, Sect. \ref{sec_sample} describes the sample. Section
\ref{sec_modeling} describes the modeling, considering first the
relevant Newtonian equations 
and the geometric setup in Sect \ref{sec_geometry}, then 
the determination of distances and the scaling of velocity
differences in Section \ref{sec_distances}, and finally presenting the
Newtonian solutions in  Section \ref{sec_solutions}.
Section \ref{sec_results} discusses the obtained results.
Section \ref{sec_conclusions} draws our conclusions. The Appendix
\ref{App_discarded} describes the discarded pairs.

\section{Method}
\label{sec_method}

Comparing the observations of WBs with Newton theory is in
principle simple, as it requires to measure the distribution of
observed velocity differences between the components of the WBs ($V_{\rm ra}$, $V_{\rm dec}$, $V_{\rm rad}$) and their separation, and to compare the results of the
observations with theoretical predictions.  In practice, the
exceedingly long orbital periods and the small velocity difference
expected in WBs make the process more complex and also set some
limitations: for instance only angular separations on the sky are known
with precision, and
the full 3D orbit cannot be recovered.  In the following we recap the
steps used in the process and how they have been applied here or in the past.

Measure the separation: thanks to Gaia, the separation between the
stars can been computed very precisely by using Gaia coordinates and
parallaxes. Several authors \citep{Pittordis+2019, Chae+2024, Chae2025}
  have argued that the
exquisite Gaia parallaxes are not precise enough to compute the
separation in 3D, because the residual error in the stellar distances
is comparable to the stars' separation. Then projected separations {\it
  s} are computed, by assuming that both stars lie at the distance of
the system, obtained by averaging the parallaxes, weighted for their
parallax error.  The uncertainty on the separation {\it s} is better
than 0.1\%. In Sect. \ref{sec_distances} we describe how we
can go beyond this approach and estimate the intrinsic separation $r$
of the systems.
 
Remove not-suitable systems: the main
  cause of contamination of WBs samples is the presence of multiple
  systems, that will heat the velocity difference between the WBs
  components. Chance alignment, though more rare in systems with
  separations below 1 parsec \citep{ElBadry2021}
  must also be considered.
As far as multiple systems are concerned, it is important to recall
that these potential intruders are close companions of one of the
stars unresolved by Gaia.  The main tools we use to determine close
companions are the variability of multiple precise RVs and the
renormalized unit weight error (RUWE) parameter of
Gaia. \citet{Kervella2022} argue that a RUWE value of around 1.0 is
expected for a single-star astrometric solution. In contrast, a value
significantly larger than 1.0, typically larger than 1.4, could
indicate a non single source or one problematic for the astrometric
solution. Given the rather good agreement between the HARPS and Gaia
RVs (see below), a very large difference (several km~s$^{-1}$ )
between HARPS and Gaia RVs also indicates the presence of a companion.
\citet{Penoyre2022} noted that RUWE values of less than 1.25 could
still be multiple systems where the inner periods are a significant
multiple of the GAIA baseline observation period (34 months for DR3).
However, the fraction of binaries that can significantly survive a
RUWE=1.25 threshold is considerable only for orbital periods of a few
decades and drops at a few percent for binaries with 100 yr period.
But we are able to identify such systems (provided they are not in
extreme face-on configurations) through multiple HARPS observations.
As an example, a $2M_\odot$ binary at a separation of 20 AU has an 
orbital period of 63 yr; observing the system with HARPS one year apart allows
one to detect the expected variations in the radial velocity
differences (more than 30 m~s$^{-1}$ in 84\% of the cases
for an eccentricity of 0.6 and random orientations). Only binary \#27
of our sample has an extremely low (0.02) radial to tangential velocity
ratio (see Sect. \ref{sec_results}).

It is not obvious how to establish a priori a firm criterium on RV
variability for considering a star a close binary.  RV variability may
in fact depend on several factors, such as: precision of the
measurements, presence of companions, rotational velocity and spectral
type \citep{Bouchy+2001}, stellar activity and pulsations.  The
typical precision of the HARPS single exposure is of a few m~s$^{-1}$,
but no correction has been adopted for long term instrumental drifts,
nor whether different versions of the pipeline have been used to
measure the RVs \citep{Barbieri+2024}. For instance, if an object has
been observed with HARPS before and after the fiber change
\citep{LoCurto+2015}, a zero point offset of about 15 m~s$^{-1}$ is
expected for solar stars. The RV variability induced by stellar
activity, either due to rotational modulation or to long term stellar
cycles, is difficult to estimate, but all our WBs are reasonably old
($>$ 1 Gyrs).  When considering the 50 m~s$^{-1}$ RV modulation
measured in the young ($\sim 8\times 10^8$yrs) Hyades stars caused by
rotational modulation \citep{Saar+1997} and that activity decreases
exponentially with age in the first two Gyrs, to reach a level
compatible to the Sun \citep{Pace+2004}, we estimate that rotational
modulation should contribute to RV variability not more than a dozen
m~s$^{-1}$ (peak to peak), as observed by HARPS in the Sun
\citep{Haywood+2016} .  Long term cycle RV variability (equivalent to
the 11-yrs solar cycle), which is relevant for observations spanning a
time baseline of years, is also poorly characterized, but the
variations of the solar RV over almost a decade have been measured to
less than 5 m~s$^{-1}$ with HARPS \citep{Lanza+2016}.  Considering all
these factors, a long term RV velocity variability of the order of 10
m~s$^{-1}$ if fully compatible with a binary system. Only larger
values may be seen as a signature of a third component.

The Gaia  RUWE parameter is also a powerful tool
to identify binaries \citep{Belokurov+2020} and different
maximum values have been suggested to limit the sample to single
stars, with \citet{Lindegren+2018} suggesting RUWE $<$1.4 for DR2 data
and \citet{Penoyre+2022} finding that RUWE = 1.25 is the most extreme
value expected from stochastic scatter of single stars.  As a
reference, our final sample has a median RUWE of 0.975 and the highest
RUWE is 1.24.

To find systems aligned by chance, we adopt chemical composition as
primary indicator: the high resolution spectra obtained for the RV
measurements were used to determine precise stellar parameters and
chemical composition \citep{Spina+2021}. In our sample (see
next section) we discarded as alignment chances those systems whose
components [Fe/H] differ by more than  3 times their associated
uncertainty in [Fe/H].

In order to minimize the contamination by
proper motion pairs, we excluded cases younger than 1 Gyrs and/or
belonging to known clusters.  Common proper motion pairs have in fact
a reasonable high probability to survive in the first Gyr of their
life \citep{Jiang+2010}.

In the absence of firm physical limits on velocity variability and RUWE
parameter to be imposed a priori, we opted to clean the original
sample in an interactive way: we used the precise HARPS RVs
to individuate stars showing large RV
variations, eliminating WBs with RV variability exceeding 100
m~s$^{-1}$ and with high RUWE ($>1.4$). We carefully looked at the
remaining stars, and in some doubtful case, we excluded stars that had
both  high RV variability and a large RUWE
parameter. When the HARPS RV measurement differs substantially from
Gaia (several km~s$^{-1}$), we also remove the star as a likely
binary. We consider our selection process as conservative (cfr. next
section and Appendix \ref{App_discarded}, where we motivate our decisions).
  
Compute the 3D velocities: the 3D velocity difference $V_{\rm tot}$
  has two components: the astrometric component (transverse or tangential 
  velocity) and the RV. The tangential velocity difference
  is usually computed by scaling the Gaia proper motions and errors
  with the mean distance of the system. In Sect.  \ref{sec_distances}
  we go beyond this simplified approach by exploiting the distance probability
  distributions delivered by Gaia. We compute velocities in the right ascension and declination direction by
  scaling the proper motions and relative errors with the assumed
  distances of the two stars separately before taking their
  differences.

  Since the two stars forming a WB have not
  the same physical parameters, we need to obtain accurate RVs,
  properly correcting the observed shifts.  To compute the absolute
  RV we used the process adopted in \citet{Leao+2019}.
  These authors showed that HARPS observed shifts, corrected for
  gravitational redshift and convective motions, agree with astrometric
  velocities to better than 40 m~s$^{-1}$. A similar method has been
  applied to the $\alpha$ Cen system by \citet{Kervella+2017} with a
  comparable accuracy.  From the observed HARPS Cross Correlation
  Function (CCF) velocities we subtracted the gravitational redshift and
  the correction for convective shift. Both quantities depend on basic
  stellar parameters. Gravitational redshift depends on the stellar
  mass and radius, while the convective shift is computed by using 3D
  atmospheric models that depend mostly on stellar effective
  temperature and gravity \citep{Ludwig2009}.  The metallicity of our
  sample stars is sufficiently close to solar, that only solar
  metallicity atmospheric models can be used.  In summary, the
  RV  of each star $i$ has been computed as
  $RV_i = V_{{\rm HARPS},i} - GR_i - V_{{\rm con},i}$,
  where $V_{{\rm HARPS},i}$ is the HARPS velocity shift measured by fitting the mask
  cross-correlation function and determined to some m~s$^{-1}$ precision,
  $GR_i$ is the gravitational redshift and
  $V_{{\rm con},i}$ is the convective (blue) shift estimated by
  cross-correlating the synthetic spectra generated with 3D models
  \citep{Ludwig2009,Leao+2019} with the same digital mask used for the
  observations. 

  As far as the associated error to $RV_i$, the systematic
  uncertainties on the gravitational redshifts cancel out when the
  velocity differences of the pairs are computed, given that masses
  and radii of the pairs are never too dissimilar and are computed
  with the same method. The uncertainty on the convective shift
  correction is not precisely known, and it is estimated of the order
  of a few tens of m~s$^{-1}$ .  The uncertainty of 40 m~s$^{-1}$
  obtained for the Hyades stars by \citet{Leao+2019} and the estimated
  error of \citet{Kervella+2017} empirically confirm this estimate.
  This factor is the main source of error in the RV measurements. The
  impact of the gravitational redshift and convective shift on the RV
  difference of the pair is in all but one WB small, but becomes relatively
  more important,  the smaller the relative velocities of the WB are.  

\section{Sample}
\label{sec_sample}

The HARPS science products available in the ESO archive contain, in
addition to the spectra, also the computed RV, together with the mask
used for the cross-correlation and other characteristics of the CCF.
An interface has been recently built to retrieve the RVs and other CCF
parameters from the catalogue of HARPS observations
\citep{Barbieri+2024}. In the past 20 years, several WB candidates
have been observed with HARPS in the framework of different programs,
either dedicated to the search of exoplanets or to study the detailed
chemical composition of WBs \citep{Spina+2021}.  The list of Spina et
al. has been the main input for the HARPS search, that has produced 44
WB candidates.
 
Most of the stars have been analyzed differentially obtaining very precise
effective temperature, gravity and abundance determinations.
\citet{Spina+2021} quote errors of 10 K or less on temperatures, 0.2
dex or less on log(g) and 0.03-0.06 dex on [Fe/H].  The tiny abundance
differences between the pairs might be the product of evolutionary
effects like diffusion, or induced by the presence of exo-planetary
systems or debris contamination \citep{Spina+2021} rather than
evidence of different birthplace.  Therefore, we have first cleaned
the sample of binaries for stars with a [Fe/H] abundance difference
larger than 3 times the [Fe/H] uncertainty in their components.
One pair was discarded with this step. 
All the final pairs have
differences in [Fe/H]  abundance of less than 0.1 dex but
two: binaries \#12 and \#21, that have the largest uncertainty in their
[Fe/H] determination.
The mean difference in [Fe/H] is -0.01 dex with 0.06 dex
RMS. We also used the estimated stellar ages and their 95\% confidence
levels (determined while computing the stellar masses and radii, see
below) to verify that the stars are older than 1 Gyr, that the age
estimates for the components of the same WB are compatible within the
errors.  In this step one pair was eliminated: with an age estimate of
700 Myr and belonging to the Hyades. No other pair could be associated
to a cluster, nor any other young star is left in the final sample.
The average difference between the ages of the pairs is 0.3 Gyr with 3
Gyr RMS.

The sample is heterogeneous in terms of number of observations and
time baseline. All stars but three have several HARPS observations,
ranging from 2 to more than 90, covering a timeline that spans from
weeks to more than 10 years.  Since the majority of the stars has
multiple HARPS observations, we are able to perform the cleaning using
RV variability.  RVs are read off the \citet{Barbieri+2024}
catalogue, inspected, and also compared with the Gaia RVs
\citep{Katz+2019}. The nature of our WBs requires some caution in
using the catalogue: some of the components of the WBs are so close on
the sky that the 3.6m telescope coordinates do not help in
discriminating which star has been observed, and it is risky to rely
only on the data header, so the observations must be 
carefully scrutinized.  In those few doubtful cases the observations (RV, width of
the CCF) have been inspected and they provide a clear information on
which star has been observed.  In one case we found an unexplained
shift of several km~s$^{-1}$ in the HARPS catalogue, that, after
direct inspection of the spectra, was not present in the original
data.  This is to emphasize that for this sample we could not use the
catalogue blindly, but a manual cleaning has been necessary.  For each
star we measured the mean RV and its RMS.  For three stars, HARPS
observations are not available (WASP-94 B, HIP47839, HD135101B):
for them we adopted the difference in RV with respect to the other
binary component as measured by Gaia, with an associated RV error of
300 m~s$^{-1}$ (see below).  For HD135101A, out of 71 observations,
two show a different RV, and CCF width. We attributed these
observations to the companion, HD135101B.
The RUWE values were finally used, as explained above.

After the cleaning process, 32 binaries (64 stars, or 73\%) out of 44 survived
the selection.  The Gaia parameters
(right ascension, declination, parallax, proper motions and RUWE
values) of the 64 stars are given in Table \ref{tab_Gaia}.  The
stellar properties (age, effective temperature, gravity, metallicity,
mass, radius, gravitational redshift, convective shift, mask and
number of HARPS observations, RV) are given in Table \ref{tab_Stars}.
Temperature, gravity and metallicity come from \citep{Spina+2021}; the
missing values are listed as $9999\pm 9999$; they are available from
\citet{Perdelwitz2024}, but we do not use them for consistency.
The gravity values given in italics are Gaia values. Age,
mass, radius and convective shifts are computed as described below.
All the binaries are limited to on-sky separations of less than 20
kAU.  The discarded pairs are described in the Appendix
\ref{App_discarded}, together with the rejection reason.

The median RV variability for the final sample is 10
m~s$^{-1}$ with 6 m~s$^{-1}$ RMS.  Some of the stars with a high RV
variability, such as HD20782 or HD106515A, are known to host massive
exo-planets, and this clearly enhances their stellar RV
variability. None of our stars show a RMS larger than 29
m~s$^{-1}$ (HD106515A).
 
Summarizing the results of our cleaning process, we can confirm that
{\it for this sample of observations} a maximum RUWE value of 1.25,
and, a long-term RV variability of 10 with 6 m~s$^{-1}$ RMS would
have been a priori good choices to select pairs composed by single
stars.
 
For most stars precise stellar parameters (T$_{\rm eff}$, log(g),
chemical composition) are given in \citep{Spina+2021}, and have been
determined by spectroscopy, applying a differential, line - to - line,
technique that maximizes the precision. These parameters have been
adopted as input to determine stellar masses, radii and ages using the
'param 1.4' interface \citep{DaSilva+2006, Rodrigues+2014,
  Rodrigues+2017}
\footnote{\url{https://stev.oapd.inaf.it/cgi-bin/param}}
and PARSEC isochrones from \citet{Bressan2012}.
For the stars without spectroscopic log(g) (listed
in italics in Table \ref{tab_Stars}) and/or metallicities from
\citet{Spina+2021}, we used the version 1.3 that accepts magnitudes
and parallaxes. For the stars with spectroscopic log(g) the two methods agree within the estimated errors. The only difference between the published data and
those used in this work is that we set the minimum T$_{\rm eff}$
uncertainty to 30 K (for most stars the published uncertainty is much
smaller), in order to allow the evolutionary code to converge, because
the algorithm may not do so for T$_{\rm eff}$ uncertainties of a few
K. The derived stellar masses and radii are used to compute the
gravitational redshifts; the convective shift corrections are computed
using the values of log(g) listed in Table \ref{tab_Stars}. They
differ slightly (within the errors) from the values derived from the
stellar masses and radii given in Table \ref{tab_Stars}.

Since what matters are the relative shifts and the relative
gravitational redshifts between the two WB components, it is more
important that the stellar parameters are on the same scale rather
than their accuracy is on an absolute scale. We estimate from the 95\%
confidence levels quoted by 'param' and the comparison between the
versions 1.3 and 1.4 that the systematic error affecting the total
masses of the binaries is at most 10\%.
  
Comparing the HARPS (not corrected) and Gaia RV we find that the zero
point differs by a median value of 68 m~s$^{-1}$, with 311 m~s$^{-1}$
RMS; when comparing our final RVs (corrected for gravitational
redshift and convective shift) with Gaia, the median zero point
difference is -37 m~s$^{-1}$ , with 300 m~s$^{-1}$ RMS.  This spread
is in line with the expected precision of the Gaia RV. The HARPS RVs
median values, not corrected for GR and convective shift, are given in
Table \ref{tab_Stars}.

\section{Modeling}
\label{sec_modeling}
 \subsection{Newtonian orbits and Geometry}
 \label{sec_geometry}

\begin{figure}
  \includegraphics[width=\linewidth]{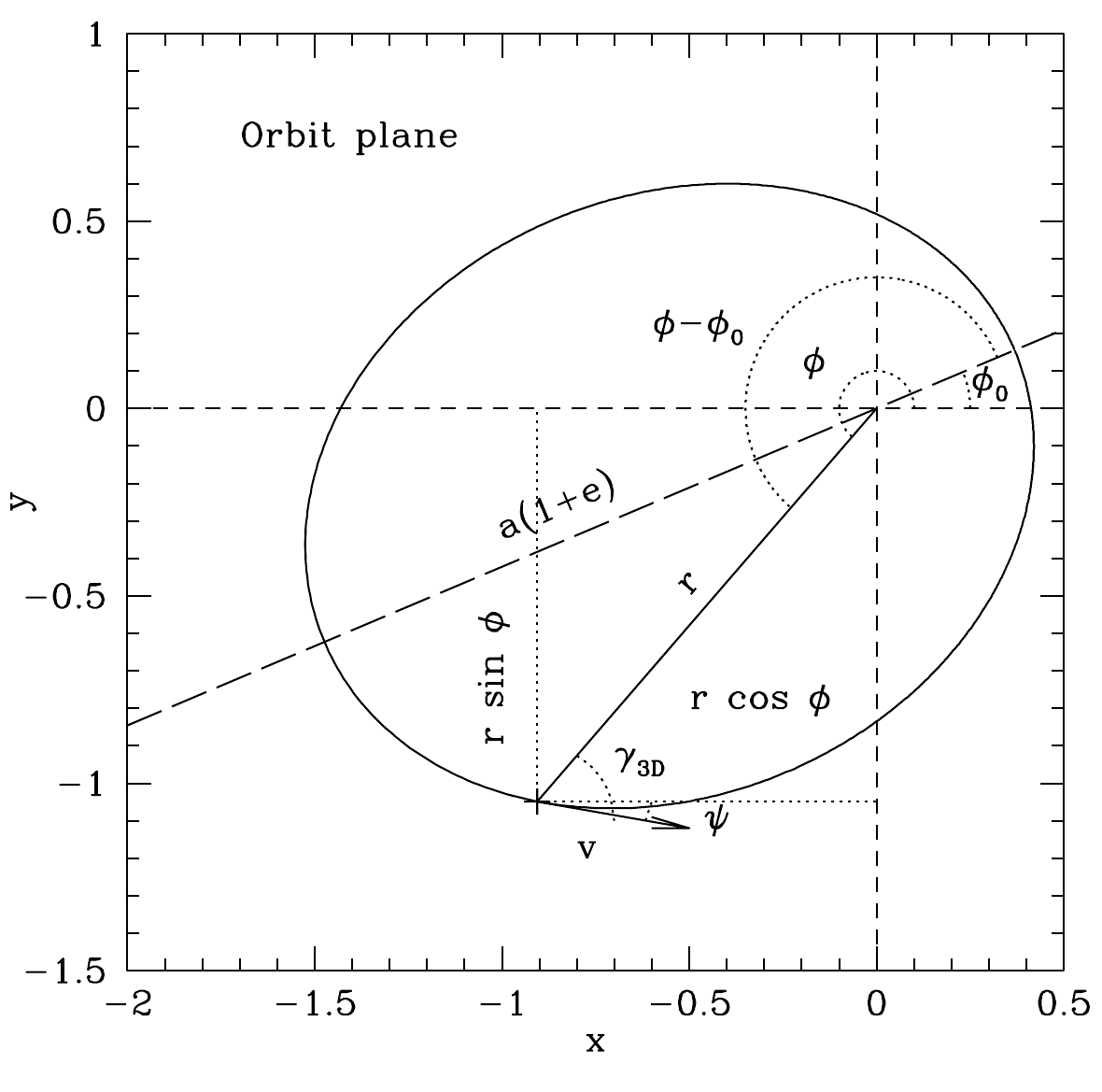}
  \caption{The plane of the orbit. The axis x is defined by the intersection of the plane of the orbit and the plane of the sky. The periastron is at
an angle $\phi_0$ from the $x$ axis; $\phi$ is the angle between the $x$ 
axis and the position on the orbit; $\psi$ is the angle between the $x$ axis and
  the velocity vector $v$. The angle between the position and velocity vectors is $\gamma_{\rm 3D}$.}
  \label{fig_orbitplane}
\end{figure}

\begin{figure}
  \includegraphics[width=\linewidth]{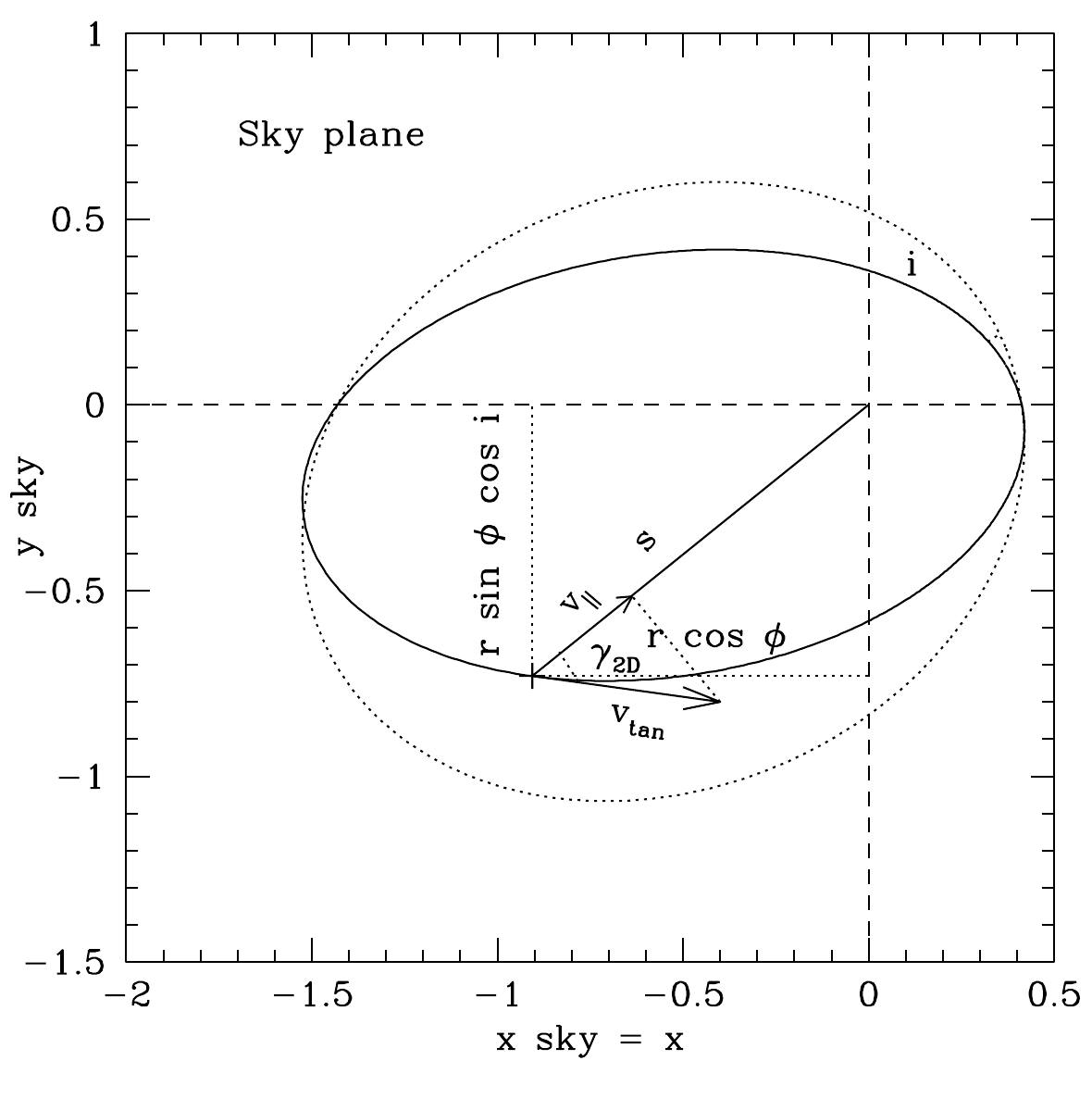}  
  \caption{The orbit projection on the sky. The angle between the position and velocity vectors on the sky is $\gamma_{\rm 2D}$. }
  \label{fig_skyplane}
\end{figure}

We define the geometry and the set of equations used to reproduce the
measured velocities assuming Newtonian dynamics, similar to the approach of
\citet{Chae+2023}. Here we adopt lower case symbols to label velocities and angles; upper case symbols are used for the quantities determined in
Sect. \ref{sec_distances}. Figure \ref{fig_orbitplane} shows the plane of
the orbit.
The reduced orbit is an ellipse with eccentricity $e$. The distance
$r$ to its focus in terms of the semi-major axis $a$ depends on the
actual position (or phase) on the orbit given by the angle $\phi-\phi_0$:
\begin{equation}
\frac{r}{a}=\frac{1-e^2}{1+e \cos(\phi-\phi_0)};
\label{eq_ra}
\end{equation}
$r/a$ can assume values between $1-e$ (pericenter for
$\phi-\phi_0=0$) and $1+e$ (apocenter for $\phi-\phi_0=\pi$):
\begin{equation}
  1-e\le r/a\le 1+e;
  \label{eq_raminmax}
\end{equation}
$r/a$ is equal to 1 when $\cos (\phi-\phi_0)=-e$.
The relative velocity of the binary expected from Newtonian dynamics is:
\begin{equation}
\label{eq_vNewton}
v(r) = \sqrt{\frac{GM}{r}(2-\frac{r}{a})}=v_{\rm N}  \sqrt{2-\frac{r}{a}},
\end{equation}
where $M=M_1+M_2$, $a$ is the semi-major axis of the orbit, $e$ the
eccentricity of the orbit, $v_{\rm N}=\sqrt{\frac{GM}{r}}$ is the
velocity expected for $e=0$ and the acceleration is $A_{\rm N}=GM/r^2$.
The period of the orbit is
$T^2_{\rm N}=\frac{4  \pi^2}{GM}a^3$. The fractional time $\Delta t/T_{\rm N}$
spent between phase $\phi_1$ and $\phi_2$ is given by:
\begin{equation}
\Delta t/T_{\rm N}=\int_{\phi_{1}}^{\phi_{2}}\frac{d\phi'}{(1+e\cos \phi')^2}/\int_0^{2\pi}\frac{d\phi'}{(1+e\cos \phi')^2}.
\label{eq_tT}
\end{equation}
Fig. \ref{fig_skyplane} shows the projection on the sky. The angle
between the plane of the orbit and the plane of the sky is the
inclination angle $i$.  The true separation $r$ of the binaries given
its projected separation $s$ on the sky is:
\begin{equation}
r=\frac{s}{\sqrt{\cos^2 \phi+\cos^2 i \sin^2 \phi}},
\label{eq_rs}
\end{equation}
where  $r \ge s$, $\phi$ is the angle between the axis $x$ defined by the
intersection of the plane of the orbit and the plane of the sky and
the position on the orbit. The periastron is at an angle $\phi_0$ from
the $x$ axis. When $i=0$ the orbit is face-on and
$r=s$. When $i=\pi/2$ the orbit is edge-on and $s/r=\cos \phi$; when
$\phi=\pi/2$ or $\phi=3\pi/2$ the position of the orbit is orthogonal
to the intersection of the plane of the orbit and the plane of the sky.
Eq. \ref{eq_rs} can be rewritten to express
$\phi$ as a function of $s/r$ and $i$:
\begin{equation}
\cos^2 \phi = \frac{(s/r)^2-\cos^2 i}{\sin^2 i},
\label{eq_rsnew}
\end{equation}
from which one finds that the minimum possible inclination angle has
$\cos i_{\rm min}=s/r$.

The component $v_{\rm rad}$ of the relative velocity $v$ along the line of sight
is given by:
\begin{equation}
  \label{eq_vrad}
v_{\rm rad}=v\sqrt{1-\cos^2\psi-\cos^2 i \sin^2 \psi}=v\sin\psi\sin i,
\end{equation}
where 
\begin{equation}
  \label{eq_psi}
\psi=\arctan\left(-\frac{\cos \phi + e \cos \phi_0}{\sin \phi+e\sin\phi_0}\right).
\end{equation}
Since $\sin \psi=v_{\rm rad}/v/\sin i\le 1$, this implies that the
inclination angle has to be larger than $i_{\rm min}=\arcsin v_{\rm rad}/v$.

The component $v_{\rm tan}$ on the sky is then
$v_{\rm tan}=\sqrt{v^2-v_{\rm rad}^2}$. The ratio:
\begin{equation}
  \label{eq_vradvtan}
  \frac{v_{\rm rad}}{v_{\rm tan}}=\frac{\sin\psi\sin i}{\sqrt{1-\sin^2\psi\sin^2 i}}
\end{equation}
is small for nearly face-on orbits with $i\approx 0^\circ$ and is approximately
equal to $\tan \psi$ for nearly edge-on orbits with $i\approx 90^\circ$.
In this case values of $\psi\approx 0$ also produce  
$v_{rad}/v_{tan}\approx 0$.
    
The angle $\gamma_{\rm 2D}$
between the position vector projected on the sky and the velocity
vector projected on the sky is given by:
\begin{equation}
  \label{eq_gamma2d}
  \cos \gamma_{\rm 2D}=\frac{e\cos^2i\cos\phi_0\sin\phi-\cos\phi(e\sin\phi_0+\sin^2i\sin\phi)}{d\sqrt{\cos^2\phi+\cos^2 i\sin^2\phi}}
  \end{equation}
where 
\begin{equation}
d=\sqrt{\cos^2i(e\cos\phi_0+\cos\phi)^2+(e\sin\phi_0+\sin\phi)^2},
\end{equation}
see Eq. A1 of \citet{Hwang2022}.
For face-on ($i=0$) orbits $\gamma_{\rm 2D}=\gamma_{\rm 3D}$ (see Eq. \ref{eq_gamma3d});
for edge-on ($i=90^\circ$) orbits, $\gamma_{\rm 2D}$ can be 0 or $180^\circ$.
The velocities parallel and orthogonal to the position vector
projected on the sky are:

\begin{equation}
\label{eq_vparperp}
  v_\parallel=v_{\rm tan}\cos\gamma_{\rm 2D}; v_\perp=v_{\rm tan}\sin\gamma_{\rm 2D},
\end{equation}
respectively.
Finally, the angle $\gamma_{\rm 3D}$
between the (tri-dimensional) position and velocity
vectors is given by:

\begin{equation}
  \label{eq_gamma3d}
  \cos \gamma_{\rm 3D}=\frac{e\sin(\phi-\phi_0)}{\sqrt{1+e^2+2 e \cos(\phi-\phi_0)}}.
  \end{equation}
The maximum value that $\cos \gamma_{3D}$ can reach is $e$ at
$\cos(\phi-\phi_0)=-e$. Both
$\cos \gamma_{\rm 2D}$ and $\cos \gamma_{\rm 3D}$ change sign when the signs of
both angles $\phi_0$ and $\phi$ are changed.

\subsection{Distances and velocities}
\label{sec_distances}

\begin{figure}
  \includegraphics[width=\linewidth]{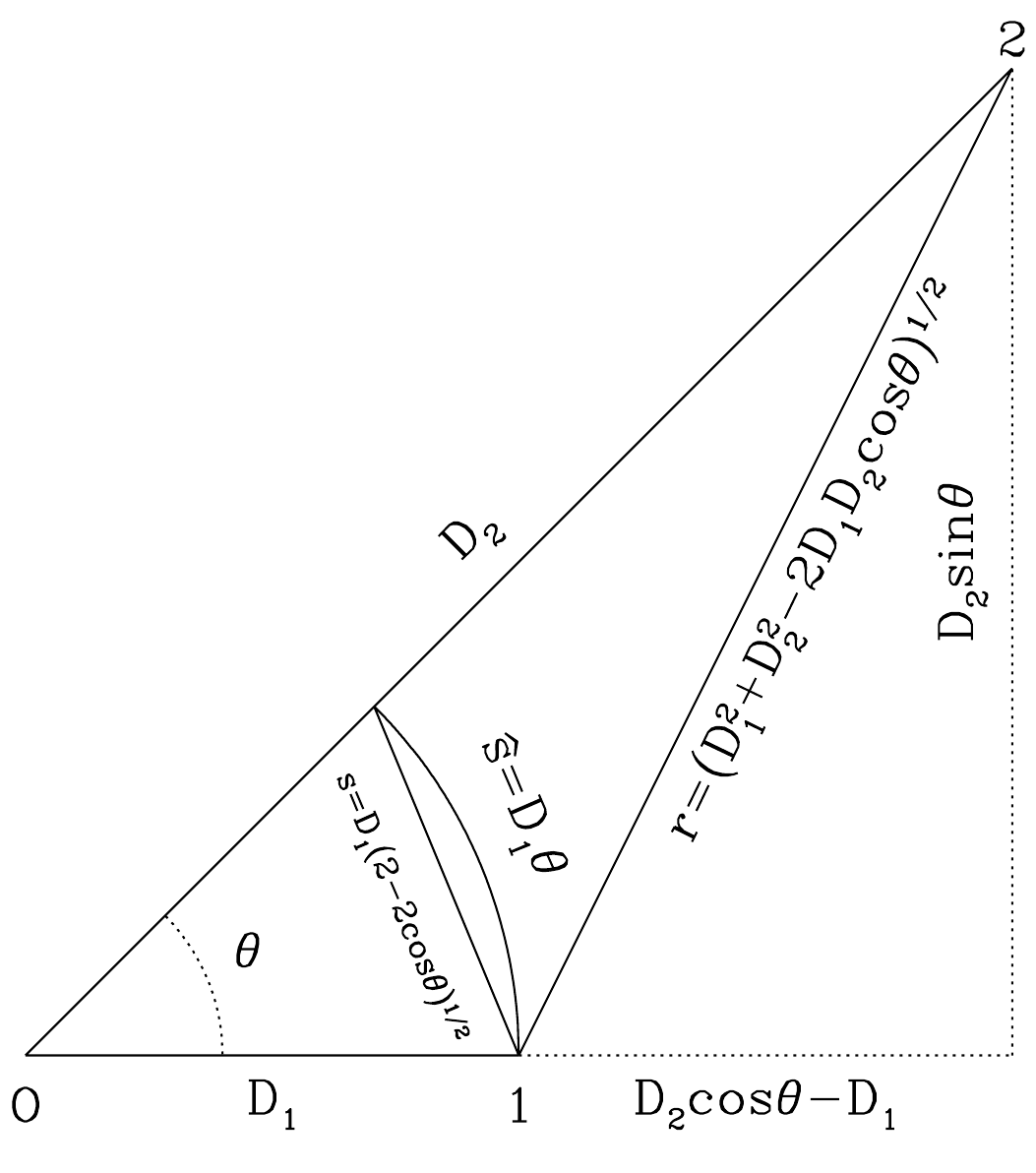}
  \caption{The geometry of the binary system from the observer's point of view.}
  \label{fig_geometry}
\end{figure}

Figure \ref{fig_geometry} shows the geometric configuration of the
binary system from the observer's point of view, where we have assumed
that star 1 (with mass $M_1$) is at a distance $D_1$ and star 2 (with
mass $M_2$) at a distance $D_2$ with $D_1<D_2$. For our sample of
binaries the angle $\theta$ between the directions of the two stars of
the pair is at most 5.4 arcmin, or 0.00157 radians, with a median
value of 18.9 arcsec, or $10^{-4}$ radians (see Table \ref{tab_chi}).
This allows us to set the chord $s$ to the length of the arc $\hat{s}$ and
simplify the projected distance on the sky of the binary to
\begin{equation}
\label{eq_s}
  s= min(D_1,D_2)\theta
\end{equation}
 and the true distance between the stars:
\begin{equation}
  \label{eq_r}
  r=\sqrt{(D_2-D_1)^2+D_1D_2\theta^2}.
\end{equation}  

The distances $D_1$ and $D_2$ follow approximately Gaussian
distributions with means $\overline{D_1}=1/p_1$ pc and
$\overline{D_2}=1/p_2$ and dispersions $\delta
D_1=\overline{D_1}\delta p_1/p_1$ and $\delta D_2={\overline
  D_2}\delta p_2/p_2$, where $p_1$ and $p_2$ are the measured
parallaxes and their respective errors $\delta p_1$ and $\delta p_2$
(see Table \ref{tab_Gaia}).  Then $s$ and $r$ are given by the
Eqs. \ref{eq_s} and \ref{eq_r}. We define $D_{\rm mean}=(D_1+D_2)/2$,
$\overline{D}_{\rm mean}=(\overline{D_1}+\overline{D_2})/2$ and
$s_{\rm mean}=\theta D_{\rm mean}$, $\bar{s}_{\rm mean}=\theta
\bar{D}_{\rm mean}$. We compute $\bar{s}$ inserting
$min(\overline{D_1},\overline{D_2})$ in Eq. \ref{eq_s} and $\bar{r}$
inserting $\overline{D_1}$ and $\overline{D_2}$ in Eq.\ref{eq_r}. We
show in Sect. \ref{sec_results} that for our sample $\bar{r}$ overestimates
$r$ by more than one order of magnitude.

The relative velocities of the binary in the right
ascension and declination directions are derived by scaling the proper
motions of the pair $(pm\alpha_1,pm\delta_1)$ and
$(pm\alpha_2,pm\delta_2)$ with the distances:
\begin{equation}
\label{eq_Vra}
  V_{\rm ra}=D_2pm\alpha_{2}-D_1pm\alpha_1,
\end{equation}
\begin{equation}
\label{eq_Vdec}
  V_{\rm dec}=D_2pm\delta_{2}-D_1pm\delta_1.
\end{equation}
In analogy with what written above, we define $\overline{V}_{\rm ra}=\overline{D}_2pm\alpha_{2}-\overline{D}_1pm\alpha_1$, $\overline{V}_{\rm dec}=\overline{D}_2pm\delta_{2}-\overline{D}_1pm\delta_1$,
$\overline{V}_{\rm ra,mean}=(pm\alpha_{2}-pm\alpha_1)\overline{D}_{\rm mean}$, $\overline{V}_{\rm dec,mean}=(pm\delta_{2}-pm\delta_1)\overline{D}_{\rm mean}$.
\citet{Chae2025,Chae+2024} use $\overline{V}_{\rm ra,mean}$ and $\overline{V}_{\rm dec,mean}$ in their analysis. 

The errors on $V_{\rm ra}$ and $V_{\rm dec}$ are in most of the cases less than 0.1\%:
\begin{equation}
\label{eq_errVra}
  \delta V_{\rm ra}=\sqrt{D_2^2\delta pm\alpha_{2}^2+D_1^2\delta pm\alpha_1^2},
\end{equation}
\begin{equation}
\label{eq_errVdec}
  \delta V_{\rm dec}=\sqrt{D_2^2\delta pm\delta_{2}^2+D_1^2\delta pm\delta_1^2}.
\end{equation}
The relative velocity of the binary projected on the plane
of the sky is:
\begin{equation}
  \label{eq_Vtan}
  V_{\rm tan}=\sqrt{V_{\rm ra}^2+V_{\rm dec}^2}.
\end{equation}

The angle between the vector connecting the pair and the vector of
velocities on sky is:
\begin{equation}
  \label{eq_Gamma2D}
  \cos \Gamma_{\rm 2D}=\frac{(\alpha_1-\alpha_2)\cos \bar{\delta} V_{\rm ra}+(\delta_1-\delta_2)V_{\rm dec}}
  {\sqrt{(\alpha_1-\alpha_2)^2\cos\bar{\delta}^2+(\delta_1-\delta_2)^2}V_{\rm tan}},
\end{equation}
where $(\alpha_1,\delta_1)$ and $(\alpha_2,\delta_2)$ are the RA and
DEC coordinates of the pair and $\bar{\delta}=(\delta_1+\delta_2)/2$.
Then, the components of $V_{\rm tan}$ parallel and orthogonal to the separation
vector are
\begin{equation}
  \label{eq_Vparperp}
  V_\parallel=V_{\rm tan}\cos \Gamma_{\rm 2D}; V_\perp=V_{\rm tan}\sin \Gamma_{\rm 2D}.
\end{equation}  
We compute the errors $\delta V_\parallel$, $\delta V_\perp$, $\delta V_{\rm tan}$
and $\delta \Gamma_{\rm 2D}$ by generating
100 Monte Carlo samples of $V_{\rm ra}$ and $V_{\rm dec}$ and computing the RMS
of the generated samples of $V_\parallel$, $V_\perp$, $V_{\rm tan}$
and $\Gamma_{\rm 2D}$. The errors reported in Table \ref{tab_binaries} are on
average 2\% for $V_{\rm ra}$ and $V_{\rm dec}$.
The error on the angle $\Gamma_{\rm 2D}$ is on average 0.3 degrees.

The (distance independent) component of the relative velocity of the binary
along the line of sight is: 
\begin{equation}
\label{eq_Vrad}
  V_{\rm rad}=RV_{2}-RV_{1}
\end{equation}
with an error
\begin{equation}
\label{eq_errVrad}
  \delta V_{\rm rad}=\sqrt{\delta RV_{2}^2+\delta RV_{1}^2+\delta V_{\rm cor}^2},
\end{equation}
where $\delta RV_{\rm 1}$ and $\delta RV_{\rm 2}$ are the errors on the
measured RVs and $\delta V_{\rm cor}=40$ m~s$^{-1}$ takes
into account the uncertainties in the gravitational redshift and
convection motion corrections.  On average, $\delta V_{\rm rad}$ is on the
order of 50 m~s$^{-1}$, except for the three pairs (\#13, \#15, \#28) where
no HARPS measurements for one of the binaries are
available. Therefore, $\delta V_{\rm rad}$ is one order of magnitude
larger than $\delta V_{\rm tan}$. 

We reverse the sign of the coordinate differences, $V_{\rm rad}$, $V_{\rm ra}$,
and $V_{\rm dec}$ if $D_1>D_2$, which does not change the sign of
$\Gamma_{\rm 2D}$.

The total relative velocity of the binary is: 
\begin{equation}
  \label{eq_Vtot}
  V_{\rm tot}=\sqrt{V_{\rm tan}^2+V_{\rm rad}^2},
\end{equation}
from which we compute the ratio $R=V_{\rm rad}/V_{\rm tot}$. Its error $\delta R$
is determined by generating
100 Monte Carlo samples of $R$ and computing their RMS. Finally,
the angle between the (tri-dimensional) vector connecting the pair and the
vector of
velocity difference is:
\begin{equation}
  \label{eq_Gamma3D}
  \cos \Gamma_{\rm 3D}=\frac{sV_{\rm tan}\cos\Gamma_{\rm 2D}\pm\sqrt{r^2-s^2}V_{\rm rad}}{rV_{\rm tot}},
\end{equation}
where the $+$ applies if $D_2>D_1$ and the $-$ if $D_2<D_1$. We determine the error on $\Gamma_{\rm 3D}$ by generating
100 Monte Carlo samples of $\Gamma_{\rm 3D}$ and computing their RMS.
Since $\delta V_{\rm tan}$ is smaller than $\delta V_{\rm rad}$, the angle $\Gamma_{\rm 2D}$ is (much)
better determined than $\Gamma_{\rm 3D}$. Indeed, the error on the angle $\Gamma_{\rm 3D}$ is on average 4 degrees.

\subsection{Best-fitting Newtonian solutions}
\label{sec_solutions}

Given the small errors affecting both measured velocities and masses
of our WBs, we expect to reduce the uncertainty on the 3D separations
of the binaries by searching for Newtonian orbits that reproduce the
observed relative positions and velocities as much as possible.  We
sample the parameter space $D_1,D_2,i,\phi,e,\phi_0$ to find the best
matches between ($V_{\rm rad}, V_\parallel, V_\perp)$ and ($v_{\rm rad},
v_\parallel, v_\perp)$ as follows.  We start by sampling the distances
$D_1$ and $D_2$ according to their Gaussian distributions set by the
mean distances $\overline{D_1},\overline{D_2}$ and errors $\delta
D_1,\delta D_2$ derived from the parallaxes listed in Table
\ref{tab_Gaia}.  We select (if they exist) the pairs of distances that
deliver separations $r$ that allow for maximum velocities
$v_{max}=v_N\sqrt{2}$ (see Eq. \ref{eq_vNewton}) larger than the total
velocity $V_{\rm tot}$ (see Eq. \ref{eq_Vtot}), dropping picks with $\sin(\arccos(s/r))<V_{\rm rad}/V_{\rm tot}$, since $\sin i\ge V_{\rm rad}/V_{\rm tot}$, see Eq. \ref{eq_vrad}. The largest possible velocity is
achieved for equal distances $D=D_1=D_2$ and minimal separations
$r_{\rm min}=s= D\theta$. This sets the maximum allowed distance to:
\begin{equation}
  \label{eq_Dmax}
  D_{\rm max}=\frac{2GM}{\theta V_{\rm tot}^2}.
\end{equation}
For a given $V_{\rm tot}$ and $s$, large values of $r$ decrease $s/r$ 
(increasing the minimal allowed inclination),
require smaller values of $r/a$ (pushing the phase towards pericenter), and/or
larger values of eccentricity (see Eq. \ref{eq_vNewton}).

We compute a combined distance figure of merit
$\chi^2_{\rm D}=\chi^2_{{\rm D}_1}+\chi^2_{{\rm D}_2}$,
where
\begin{equation}
\label{eq_chiD}
  \chi^2_{{\rm D}_1}=\frac{(D_1-\overline{D_1})^2}{\delta D_1^2},
  \chi^2_{{\rm D}_2}=\frac{(D_2-\overline{D_2})^2}{\delta D_2^2};
\end{equation}
and consider the separations ranked by $\chi^2_{\rm D}$.
We note
that for the set of distances fulfilling this constraint and
$v_{\rm max}\ge V$ the resulting differences of $V_{\rm tot}$ are less than 10 m~s$^{-1}$
for the two branches of solutions $D_1<D_2$ and vice-versa. In contrast, the allowed variations in $r$ are of the order of 50\%, the ones in  $s$ a factor 10 less. The resulting allowed ratios
$s/r$ set the minimal allowed inclinations $i_{\rm min}=\arccos(s/r)$. We
sample the full range of allowed inclinations $i_{\rm min}\le i\le \pi/2$ with
1 to 3 degrees steps and determine the corresponding four allowed $\phi$
values using Eq. \ref{eq_rsnew}. The ratios $V_{\rm tot}/v_{\rm max}$ deliver the
values $r/a=2-(V_{\rm tot}/v_{\rm N})^2$ (see Eq. \ref{eq_vNewton}) that reproduce
the measured total velocity exactly and in turn are linked to the $s/r$ ratios. From these $r/a$ values we derive
the minimum allowed eccentricity $e_{\rm min}=|r/a-1|$ (see
Eq. \ref{eq_raminmax}). We sample the range of allowed eccentricities
$e_{\rm min}\le e \le 0.99$ in 0.01 steps and compute the corresponding
allowed values of $\cos(\phi-\phi_0)$ through Eq. \ref{eq_ra}, each of
which delivers two values for $\phi-\phi_0$.  For each of the allowed
values of $(i,\phi, e, \phi_0)$ we compute $v_{\rm rad}/v$ (see
Eq. \ref{eq_vrad}) and $\gamma_{\rm 2D}$ (see Eq. \ref{eq_gamma2d}) and
the quantities:
\begin{equation}
  \label{eq_allchi}
  \chi_{R}^2=\frac{(R-\sin \psi \sin i)^2}{\delta R^2};
  \chi^2_{\Gamma_{2D}}=\frac{(\Gamma_{\rm 2D}-\gamma_{\rm 2D})^2}{\delta\Gamma_{\rm 2D}^2}.
\end{equation}
We sort the $\chi^2$ quantity:
\begin{equation}
  \label{eq_chi}
  \chi^2=\chi_{R}^2+\chi_{\Gamma_{\rm 2D}}^2
\end{equation}
and we rank the values of $(i,\phi, e, \phi_0)$ according to 
$\chi^2$. Finally, we select the solutions that reproduce $\Gamma_{\rm 3D}$ best. 

\begin{figure}
  \centering
  \includegraphics[width=0.49\linewidth]{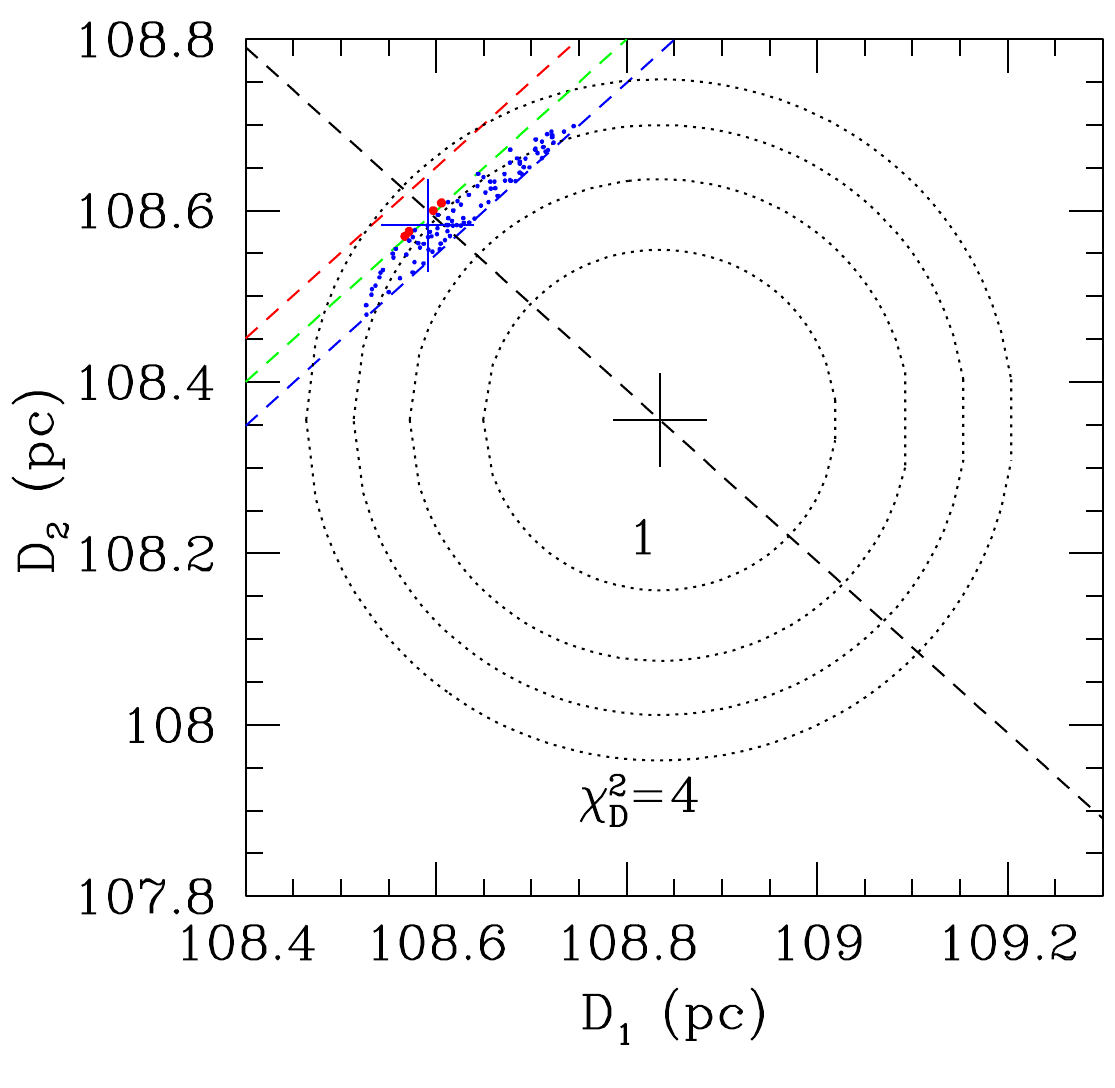}  
  \includegraphics[width=0.49\linewidth]{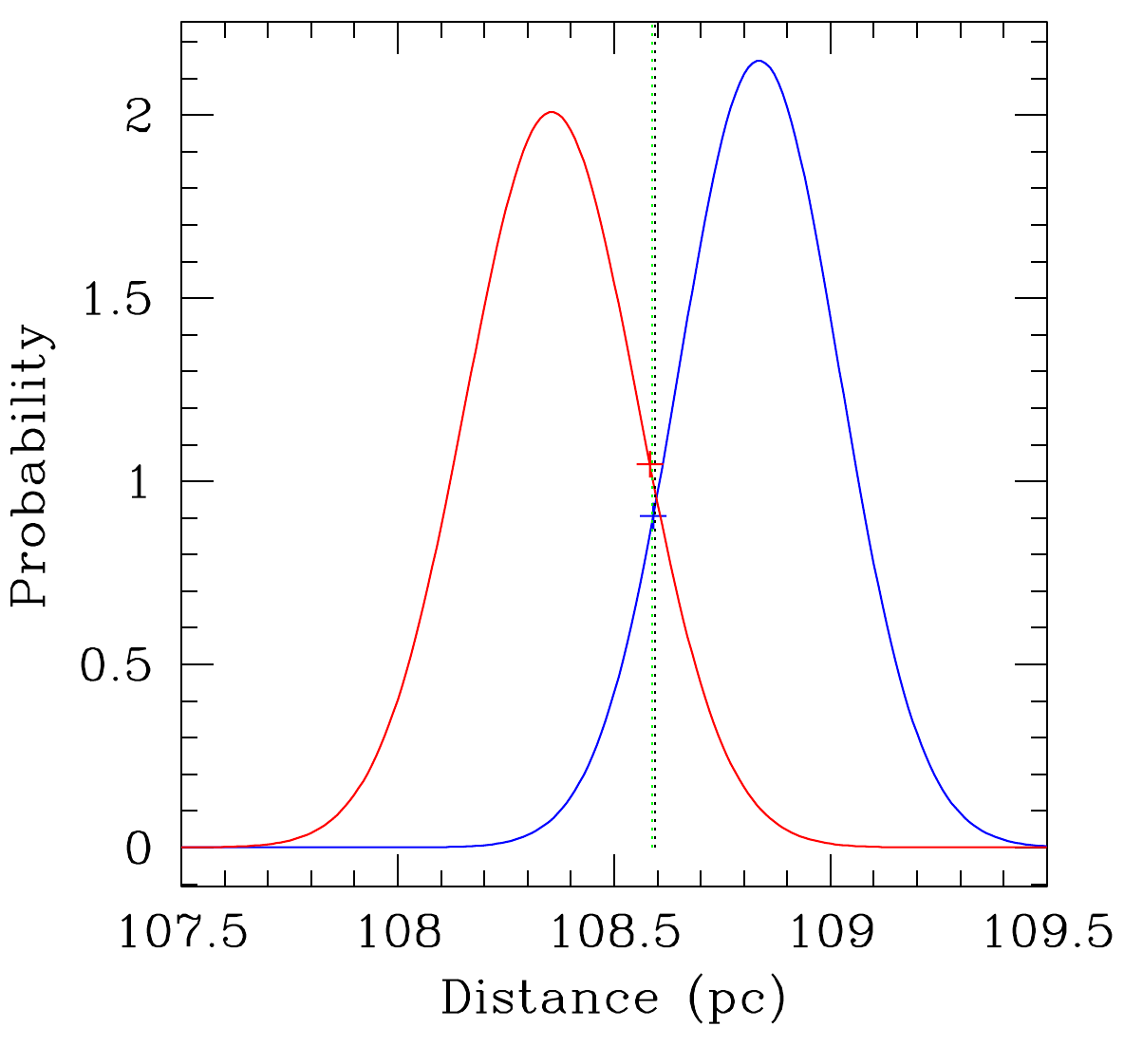}
  \caption{Distances of the binary HD4552,BD+120090. Left: the
    $\chi_{\rm D}^2=1,2,3,4$ contours (dotted lines) in the $D_1,D_2$
    plane. The black and blue crosses show the Gaia and best-fitting
    distances, respectively. The dashed black line shows
    $D_{\rm mean}=\overline{D}_{\rm mean}$, the green line shows
    $D_1=D_2$. The blue and red dashed lines show distances delivering
    $r=2GM/V_{\rm tot}^2$.  The blue and red dots show the first 100 $D_1$ and
    $D_2$ distances (ranked by $\chi^2_{\rm D}$) with
    $v_{\rm max}/V_{\rm tot}\le 1$.
      Right: the probability distributions of the distances of HD4552
      (blue) and BD+120090 (red). The crosses show the 
      best-fitting Newtonian solution. The dotted black and green
      lines show
      $\overline{D}_{\rm mean}=(\overline{D_1}+\overline{D_2})/2$ and
      $D_{\rm mean}=(D_1+D_2)/2$, respectively. }
  \label{Fig_gaussDist}
\end{figure}

\begin{figure}
\centering
\includegraphics[width=0.49\hsize]{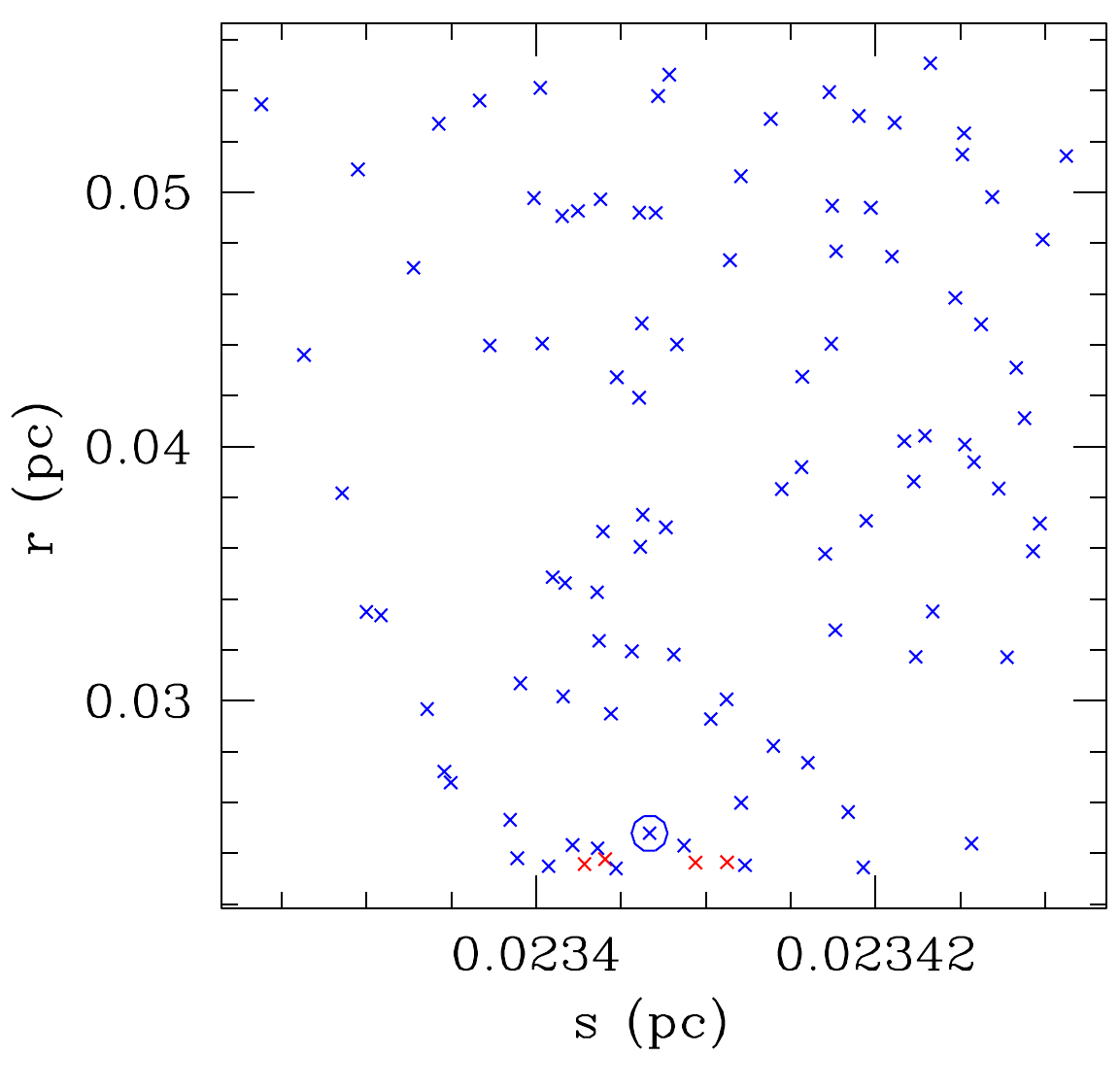}
\includegraphics[width=0.49\hsize]{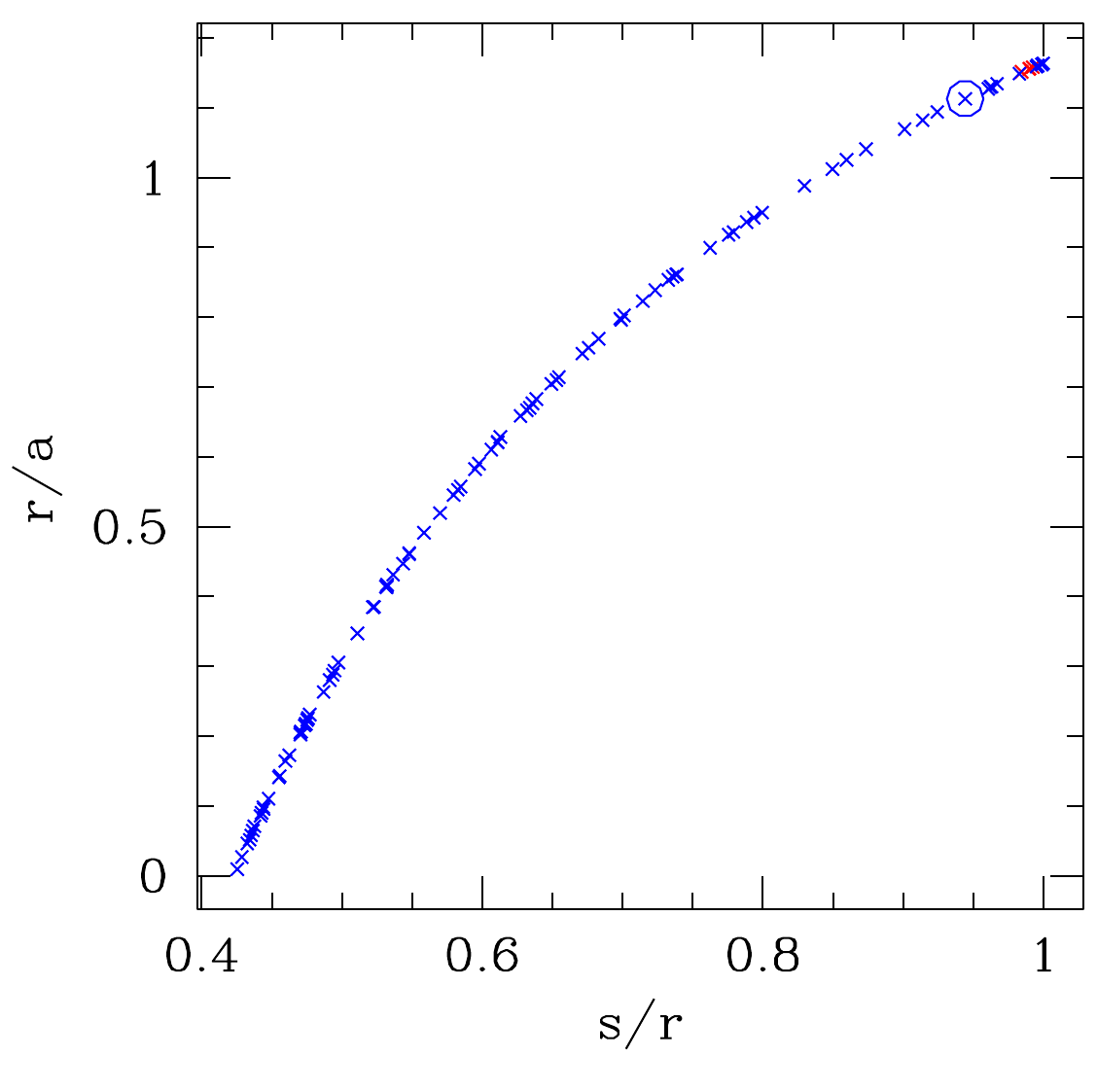}
\caption{Projected and intrinsic separations of
  the binary HD4552,BD+120090. Left: the
    distribution of projected and intrinsic separations $s$ and $r$.
    Right: $r/a$ as a function of $s/r$. Red and blue crosses show
    $D_2\ge D_1$ and $D_2< D_1$, respectively.
    The circled crosses show the best fitting solution.}
\label{Fig_Sep2DSep3D}
\end{figure}

We show an example of the minimization algorithm for the binary
HD4552,BD+120090.
Figure \ref{Fig_gaussDist}, left, presents the distribution of the best
100 distances with $v_{\rm max}/V_{\rm tot}\le 1$ ranked by $\chi^2_{\rm D}$ for
the binary \#1, HD4552,BD+120090 in the $D_1,D_2$ plane.
The allowed range of distances is narrow and translates into small
($\approx$2 m~s$^{-1}$) variations
of $V_{\rm ra}$, $V_{\rm dec}$, and $V_{\rm tan}$.
The right plot shows the probabilities of the
resulting best-fitting distances in 1D. The
best fitting solution has distances very near the average
distance of the pair and is bracket by the loci of the maximum separation
$r=2GM/V_{\rm tot}^2$ allowed by the measured total WB velocity
difference. Figure \ref{Fig_Sep2DSep3D}, left, shows that 
the separation $r$ of the binary is up to a factor 3 larger
than the separation $s$ on the sky. The right plots shows that
the value of $r/a$ needed to
reproduce the observed value of $V_{\rm tot}$ correlates with $s/r$.
In turn, the best fitting values of  $s/r$ and  $r/a$ set
the minimal allowed inclination and eccentricity (see Eqs. \ref{eq_rs} and
\ref{eq_raminmax}).

\begin{figure}
  \centering
\includegraphics[width=\hsize]{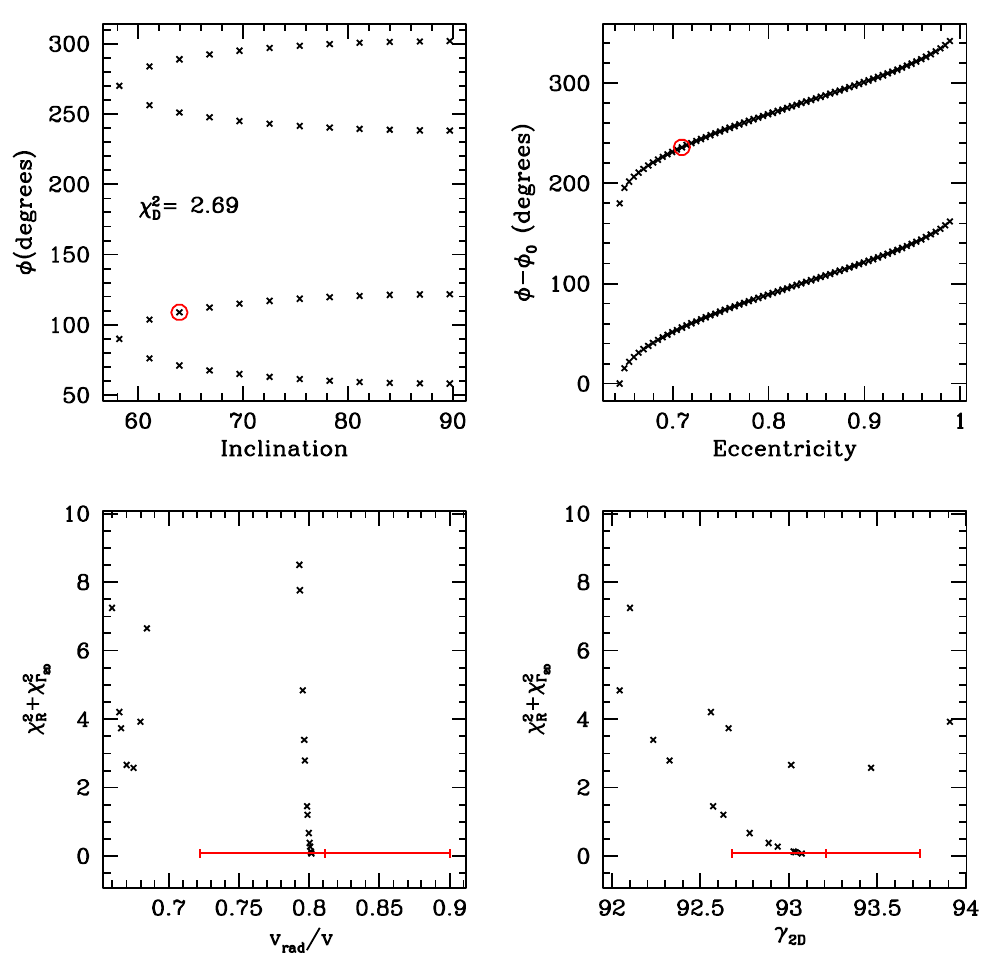}
\caption{The best-fitting Newtonian orbital solution for the
  binary HD4552,BD+120090.  Top left: the angle $\phi$ as a function
  of inclination. The red circle
  shows the selected value. Top right: the angle $\phi-\phi_0$ as a
  function of eccentricity.  The red circle
  shows the selected value. Bottom left: the black crosses show
  $\chi^2_R+\chi^2_{\Gamma_{\rm 2D}}$ as a function of $v_{\rm rad}/v=\sin
  \psi \sin i$. The red cross shows the observed value of $V_{\rm rad}/V_{\rm tot}$
  and its error (see Table \ref{tab_binaries}). Bottom right: the
  black crosses show $\chi^2_R+\chi^2_{\Gamma_{\rm 2D}}$ as a function of
  $\gamma_{\rm 2D}$. The red cross shows the value of $\Gamma_{\rm 2D}$ and
  its error (see Table \ref{tab_binaries}).}
\label{Fig_Angles}
\end{figure}

Figure \ref{Fig_Angles} illustrates how the orbit parameters
$i,\phi,e,\phi_0$ are constrained for each pair of distances selected
in Fig. \ref{Fig_Sep2DSep3D}. The inclination is sampled starting from
the minimal value set by $s/r$ and delivers four possible values for
$\phi$ (top left). The eccentricity is sampled starting from the
minimal value set by $r/a$ and delivers two values for $\phi-\phi_0$
(top right). Each combination of $i,\phi$ and $e,\phi-\phi_0$ is
considered to achieve $R=v_{\rm rad}/v$ (bottom left) and
$\Gamma_{\rm 2D}=\gamma_{\rm 2D}$
(bottom right) within the errors.
The values of $i, \phi, e, \phi-\phi_0$ that deliver
the first or, as in this case, second best of
$\chi^2=\chi^2_R+\chi^2_{\Gamma_{\rm 2D}}$ and provide the value of
$\gamma_{\rm 3D}$ nearest to $\Gamma_{\rm 3D}$ describe the best orbit
reproducing the observations. In principle, Eq. \ref{eq_gamma3d}
allows a direct determination of $\phi-\phi_0$ as a function of $e$
for a measured value of $\Gamma_{\rm 3D}$. However, the larger
observational errors affecting $\Gamma_{\rm 3D}$ make the focus on
$\Gamma_{\rm 2D}$ preferable. In the end, the pair of distances $D_1$ and
$D_2$ that deliver the optimal orbit have a $\chi^2_{\rm D}$ value slightly
larger (2.7) than the minimal one (2.5). Allowing a 2-3\% increase in
the total mass of the system is enough to decrease $\chi^2_{\rm D}$ to less
than 2. 
 
We list the values of the best-fitting solutions in the Tables
\ref{tab_chi}, \ref{tab_binaries}, and \ref{tab_modelBest}. The names
of the binary pairs are listed in Table \ref{tab_chi}. Together with
their angular separation, the fitted distances $D_1$ and $D_2$, and
$\chi^2_{\rm D}$.  Velocities $V_{\rm ra}$, $V_{\rm dec}$, $V_{\rm
  rad}$, $V_{\rm tan}$, $V_{\rm tot}$ and angles $\Gamma_{\rm 2D}$ and
$\Gamma_{\rm 3D}$ are given in Table \ref{tab_binaries}.  The
best-fitting quantities $s, r, a, e_{\rm min}, e, i_{\rm min}, i,
\phi-\phi_0, \phi_0, \psi, \gamma_{\rm 2D}, \gamma_{\rm 3D}$ are
listed in Table \ref{tab_modelBest}, together with the period and
acceleration of the orbit. These solutions are by no means unique: WBs
with low values of $\chi^2_{\rm D}$ (roughly one third of the sample)
can be fit with different combinations of distances. Given the
relatively small size of our current sample, it goes beyond the scope
of the paper to explore which (possibly non-Newtonian) solutions are also
viable in these cases.

For the binary \#24, HD189739+HD189760, we cannot find any distance
for which $v_{\rm max}\ge V_{\rm tot}$. We report a solution obtained by
increasing the total mass of the binary by 33\%.  Binary \#24 not only
requires this unrealistically large increase in mass, it also can be
fit only by eccentricities almost equal to 1 and orbital periods as
large as the age of the universe. We suspect that the system is not
bound. The binary is marked in red in the following plots.

\section{Results}
\label{sec_results}
  
\begin{figure}
  \includegraphics[width=\linewidth]{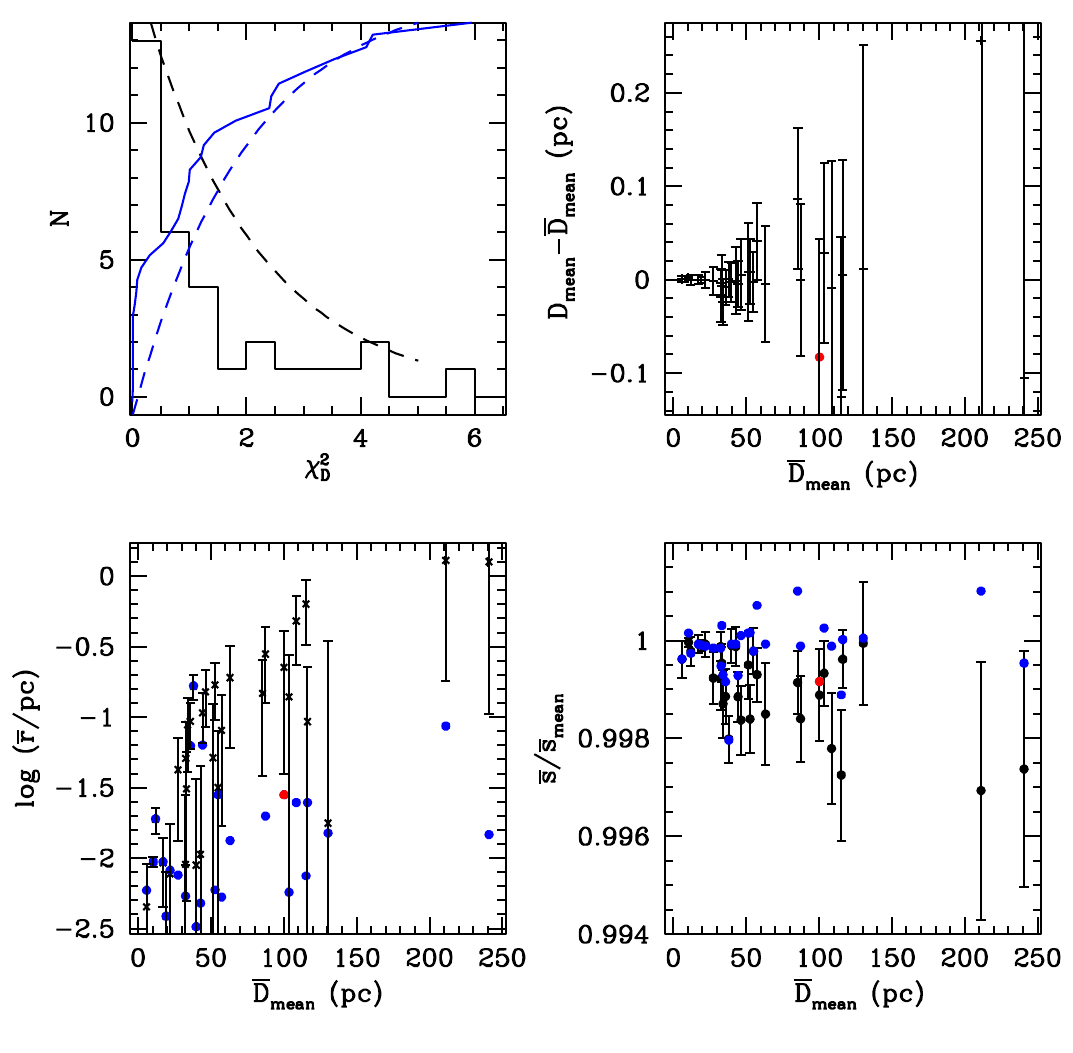}
  \caption{The distances and separations of the WB sample. Top left:
    the histogram of $\chi^2_{\rm D}$ values (black full line). The
    blue line shows the cumulative distribution of $\chi^2_{\rm D}$,
    the dashed black and blue lines show the $\chi^2$ distribution and
    its cumulative for 2 degrees of freedom. Top right: the difference
    $D_{\rm mean}-\overline{D}_{\rm mean}$ as a function of
    $\overline{D}_{\rm mean}$. The difference is zero within the
    errors. Bottom left: crosses show $\overline{r}$
    (the WB separations estimated
    using Gaia distances and Eq. \ref{eq_r}) as a function of the mean
    WB distance $\overline{D}_{\rm mean}$. The blue and red points show
    $r$, the best fitting WB separations. The largest $\overline{r}$
    are observed at the largest distances. They overestimate $r$ by an
    order of magnitude or more. Bottom right: the ratio
    $\overline{s}/\overline{s}_{\rm mean}$ as a function of
    $\overline{D}_{\rm mean}$. The blue and red points show
    $s/\overline{s}_{\rm mean}$. The WB separation on the sky
    $\overline{s}_{\rm mean}$ (estimated by scaling the angular separation
    $\theta$ with the mean WB distance), increasingly overestimates
    $\overline{s}$ (obtained by scaling $\theta$ with the minimal Gaia
    distance), as the distance to the WB increases. The best fitting
    distances deliver $s/\overline{s}_{\rm mean}\approx 1$.  }
  \label{fig_ChiD}
\end{figure}

Figure \ref{fig_ChiD} (top left) shows the histogram of $\chi^2_{\rm D}$.
Only seven pairs have $\chi^2_{\rm D}>2$ and the distribution is more peaked
at low $\chi^2_{\rm D}$ values than the $\chi^2$ distribution with 2 degrees
of freedoms. There are no obvious correlations between $\chi^2_{\rm D}$ and
$r$, $s/r$ and $r/a$. The top right plot shows that our mean distances
$D_{\rm mean}$ agree within the errors with the Gaia mean distances. The
bottom left plot shows that separations $\bar{r}$ directly inferred
from Gaia distances increase with distance and overestimate the
separations $r$ derived by fitting the measured velocities by more
than an order of magnitude. The bottom right plot shows that the WB
separation projected on the sky, estimated by scaling the angular separation
$\theta$ with the mean Gaia distance, increasingly overestimates
$\overline{s}$ (obtained by scaling $\theta$ with the minimal Gaia
distance), as the mean distance to the WB increases. Together with the
trend of increasing $\overline{r}$ with $\overline{D}_{\rm mean}$ and the large $\overline{r}/r$ ratio, this suggests that, dynamical studies of WBs samples
with similar distance
distributions and using
deprojected $\bar{s}_{\rm mean}$ run the risk of systematically
overestimating the true WB projected separation, therefore
underestimating the acceleration and velocity of the system.
This, in turn, could bias low the inferred accelerations and mimic a
low-acceleration gravitational anomaly.
The ratio $s/\overline{s}_{\rm mean}$ calculated with our best fitting distances
is approximately equal to 1 up to the largest distances and separations.

\begin{figure}
  \includegraphics[width=\linewidth]{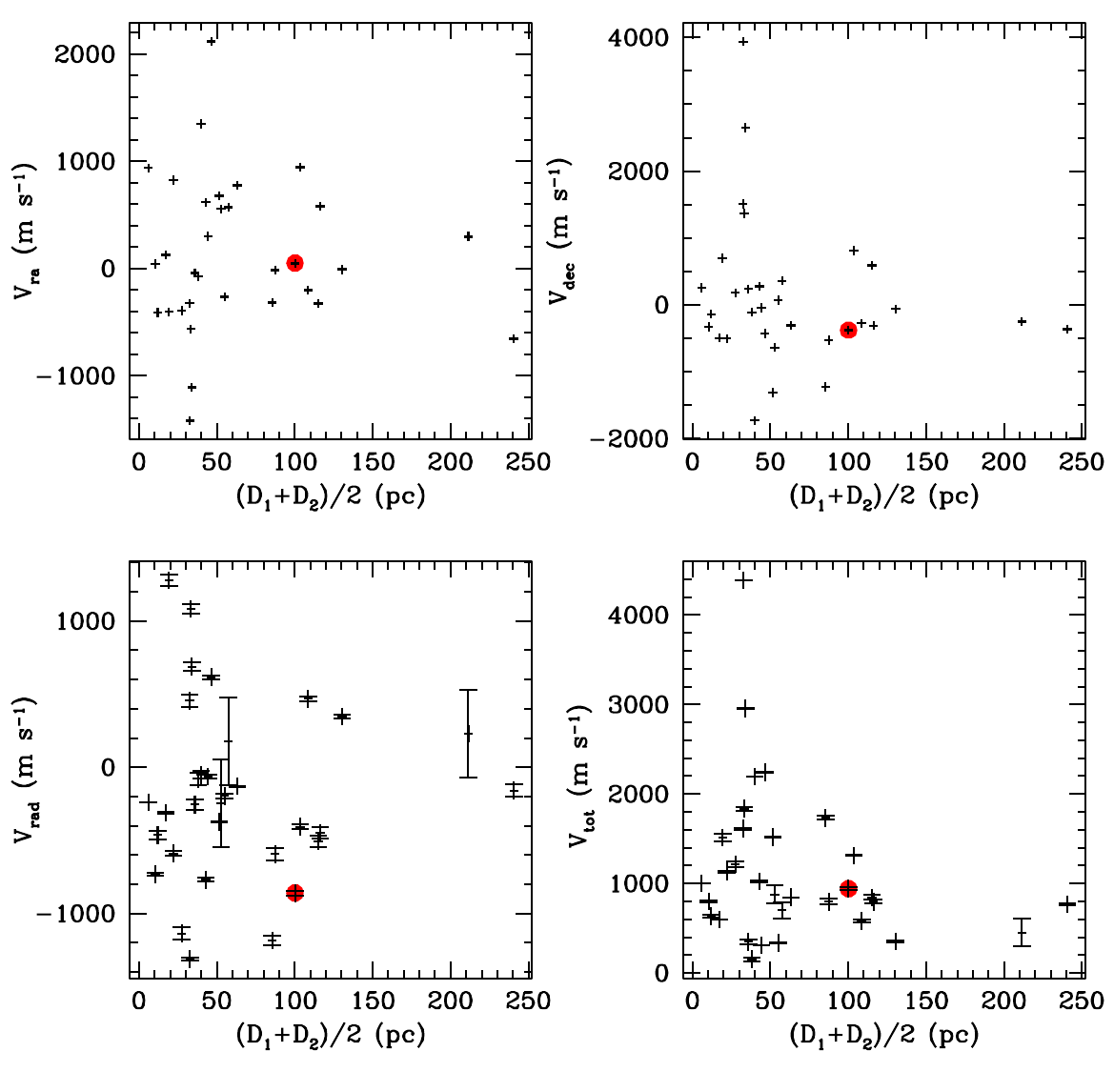}
  \caption{The velocity differences of the WBs. As a function of the mean
    distance of the pair we show $V_{\rm ra}$ (top left), 
    $V_{\rm dec}$ (top right), $V_{\rm rad}$ (bottom left) and $V_{\rm tot}$ (bottom right).}
  \label{fig_Dist}
\end{figure}

\begin{figure}
  \includegraphics[width=0.49\linewidth]{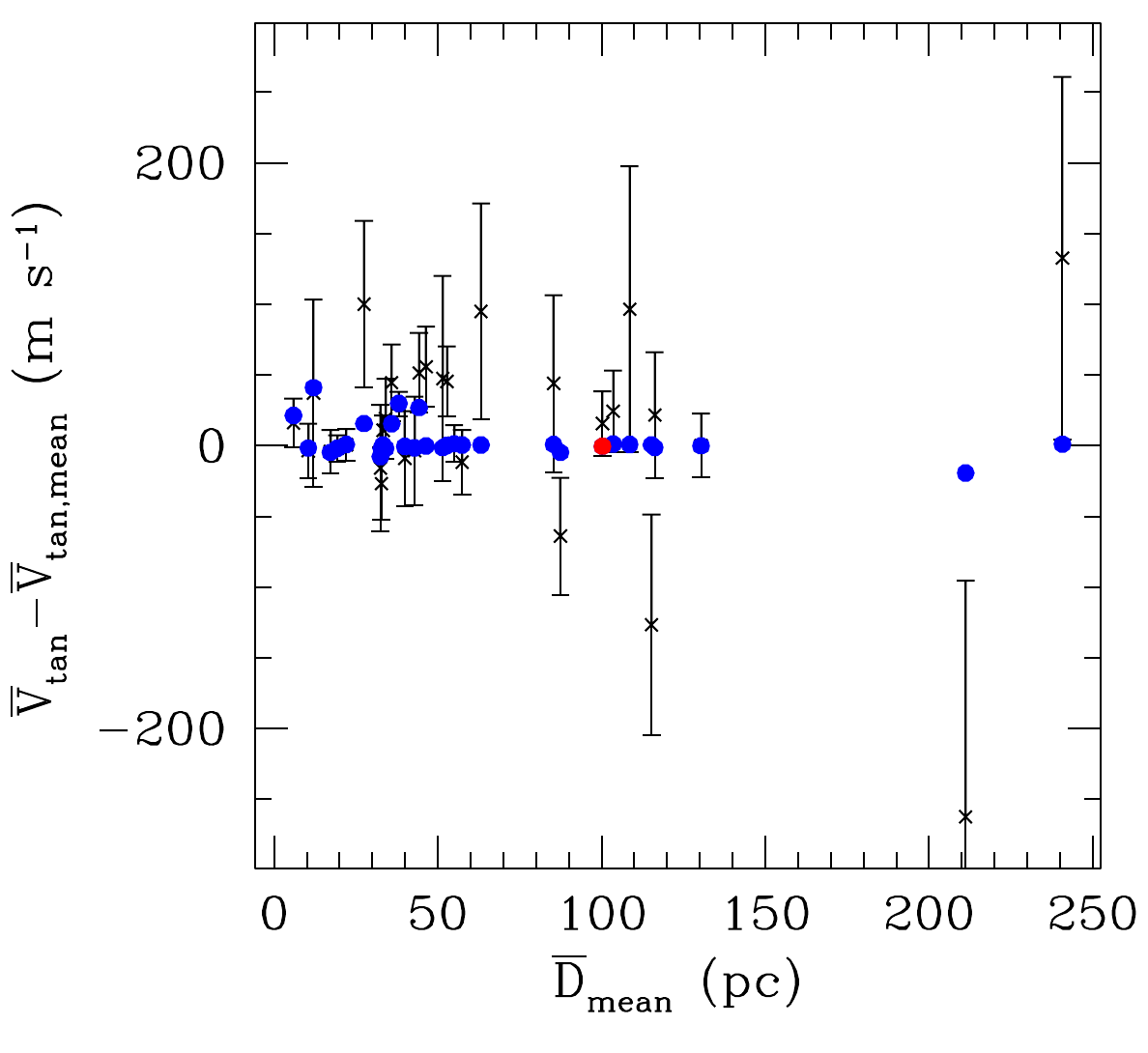}
  \includegraphics[width=0.49\linewidth]{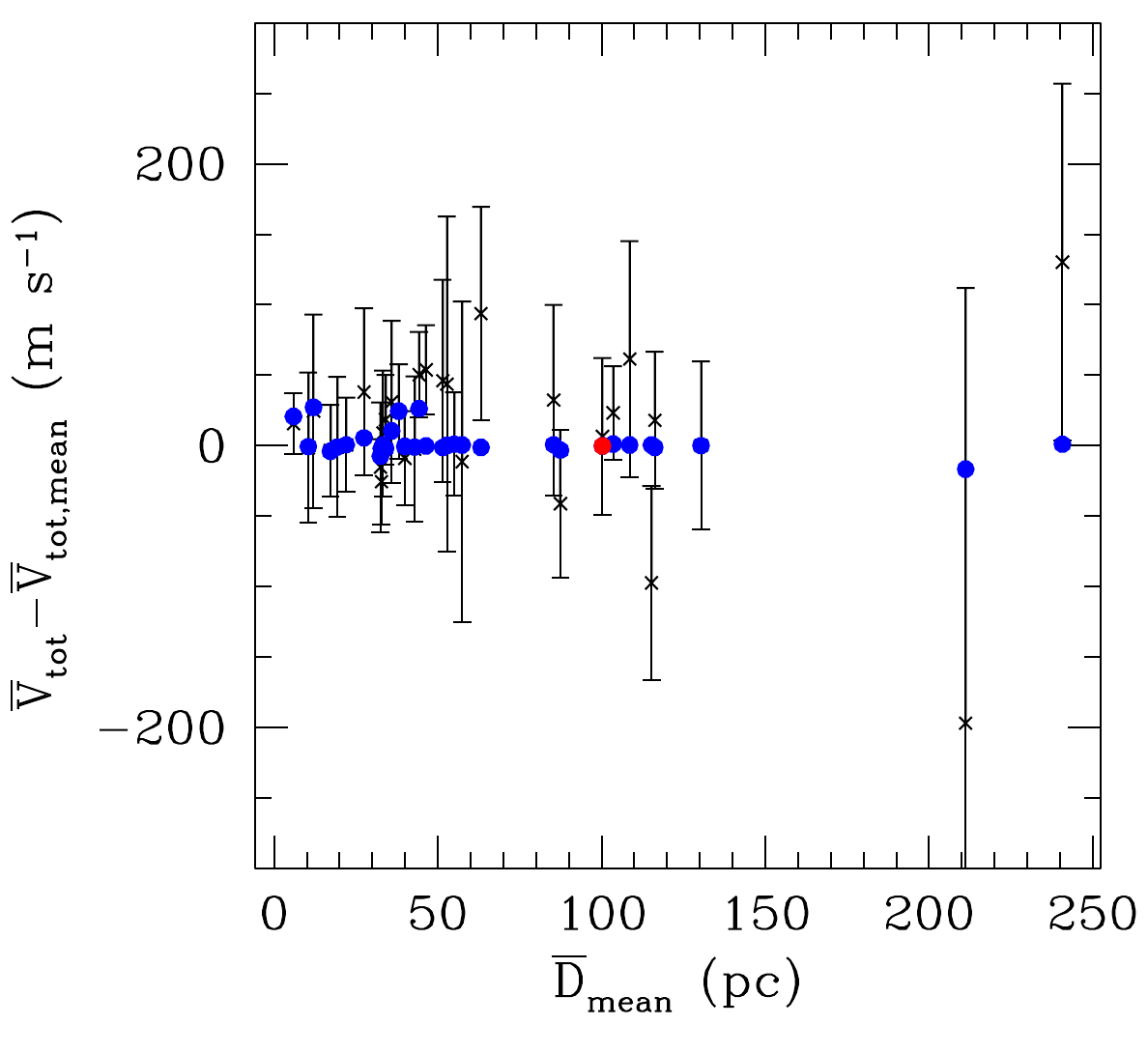}
  \caption{The influence of velocity distance scaling. The blue and red
    points show the differences between the best fitting  $V_{\rm tan}$ (left) and
$V_{\rm tot}$ (right) and $\overline{V}_{\rm tan,mean}$ and
$\overline{V}_{\rm tot,mean}$  (computed by scaling the proper motions with the
   mean Gaia distances $\overline{D}_{\rm mean}$) as a function of $\overline{D}_{\rm mean}$. The crosses show the differences between $\overline{V}_{\rm tan}$ and
$\overline{V}_{\rm tot}$ (computed by scaling the proper motions with the
Gaia distances of each star separately) and $\overline{V}_{\rm tan,mean}$ and
$\overline{V}_{\rm tot,mean}$. No systematic trends with distances are observed.
 }
  \label{fig_difVtanVtot}
\end{figure}

Figure \ref{fig_Dist} shows the velocities $V_{\rm ra}$, $V_{\rm dec}$,
$V_{\rm rad}$ and $V_{\rm tot}$ as a function of the mean distance of the
pair. Our current WB sample is clearly biased against large velocities
at large distances. We compare our velocity differences $V_{\rm tan}$ and
$V_{\rm tot}$ to the velocities differences $\overline{V}_{\rm tan,mean}$ and
$\overline{V}_{\rm tot,mean}$ computed using mean Gaia distances in
Fig.\ref{fig_difVtanVtot} (blue and red points), finding residuals
less than 40 m~s$^{-1}$ and no systematic trends with
distances. Differences between $\overline{V}_{\rm tan}$ and
$\overline{V}_{\rm tot}$ (computed scaling the proper motions with the
Gaia distances of each star separately) and $\overline{V}_{\rm tan,mean}$ and
$\overline{V}_{\rm tot,mean}$ are much larger, up to 200 m~s$^{-1}$, reflecting the
uncertainties on the Gaia distances;
still, no systematic trends with distances are observed.

\begin{figure*}
  \includegraphics[width=0.33\linewidth]{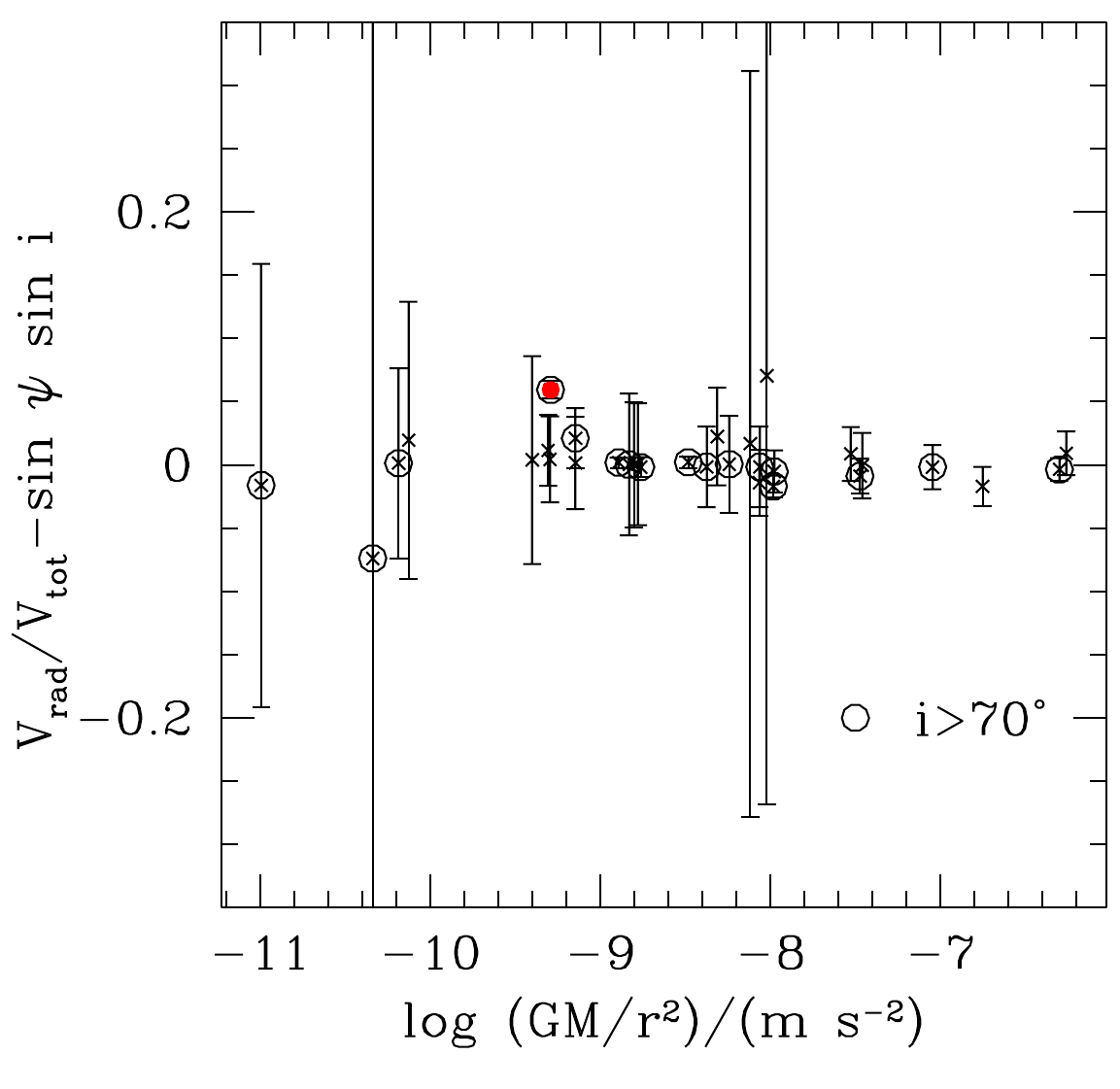}
  \includegraphics[width=0.33\linewidth]{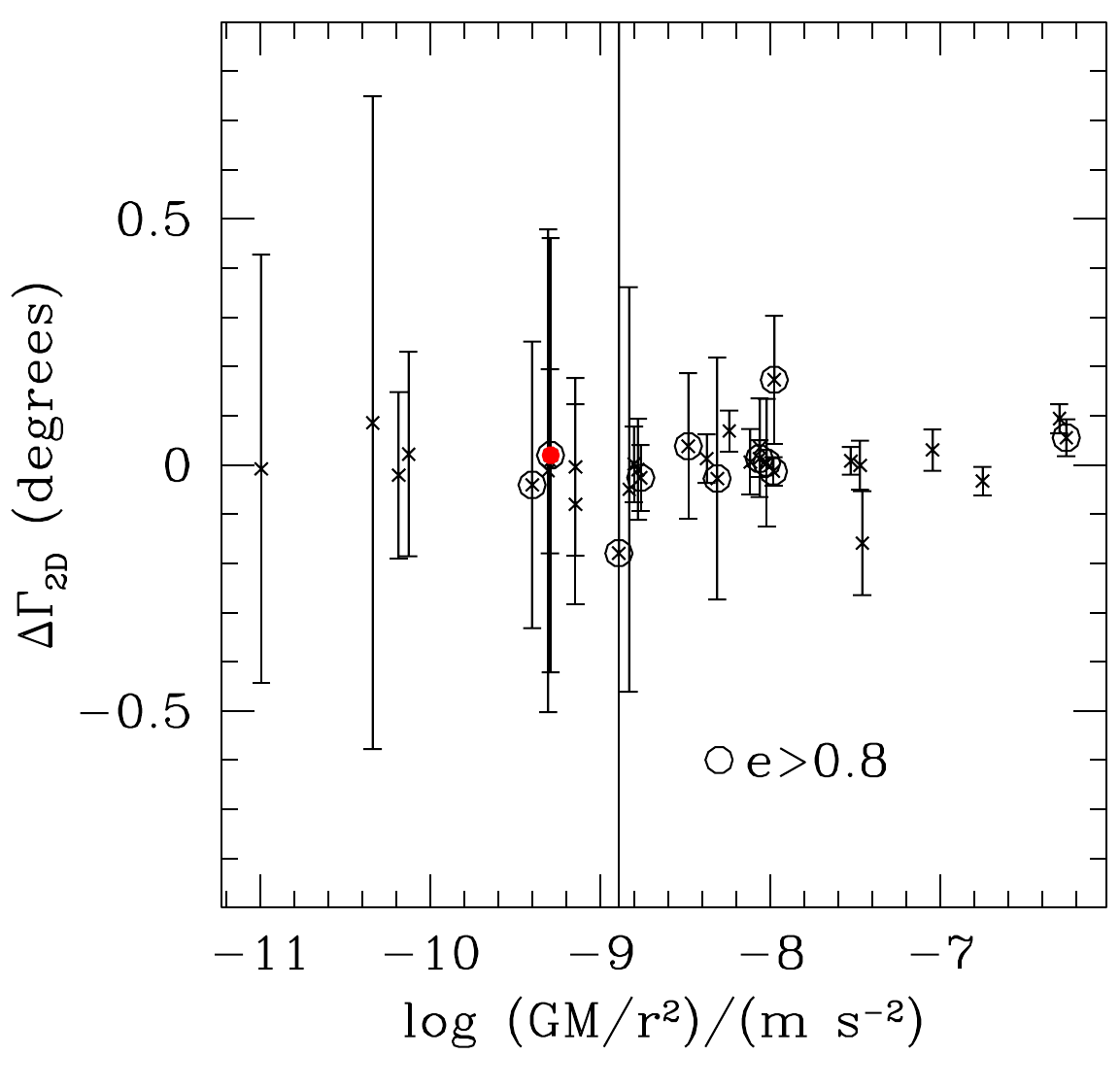}
    \includegraphics[width=0.32\linewidth]{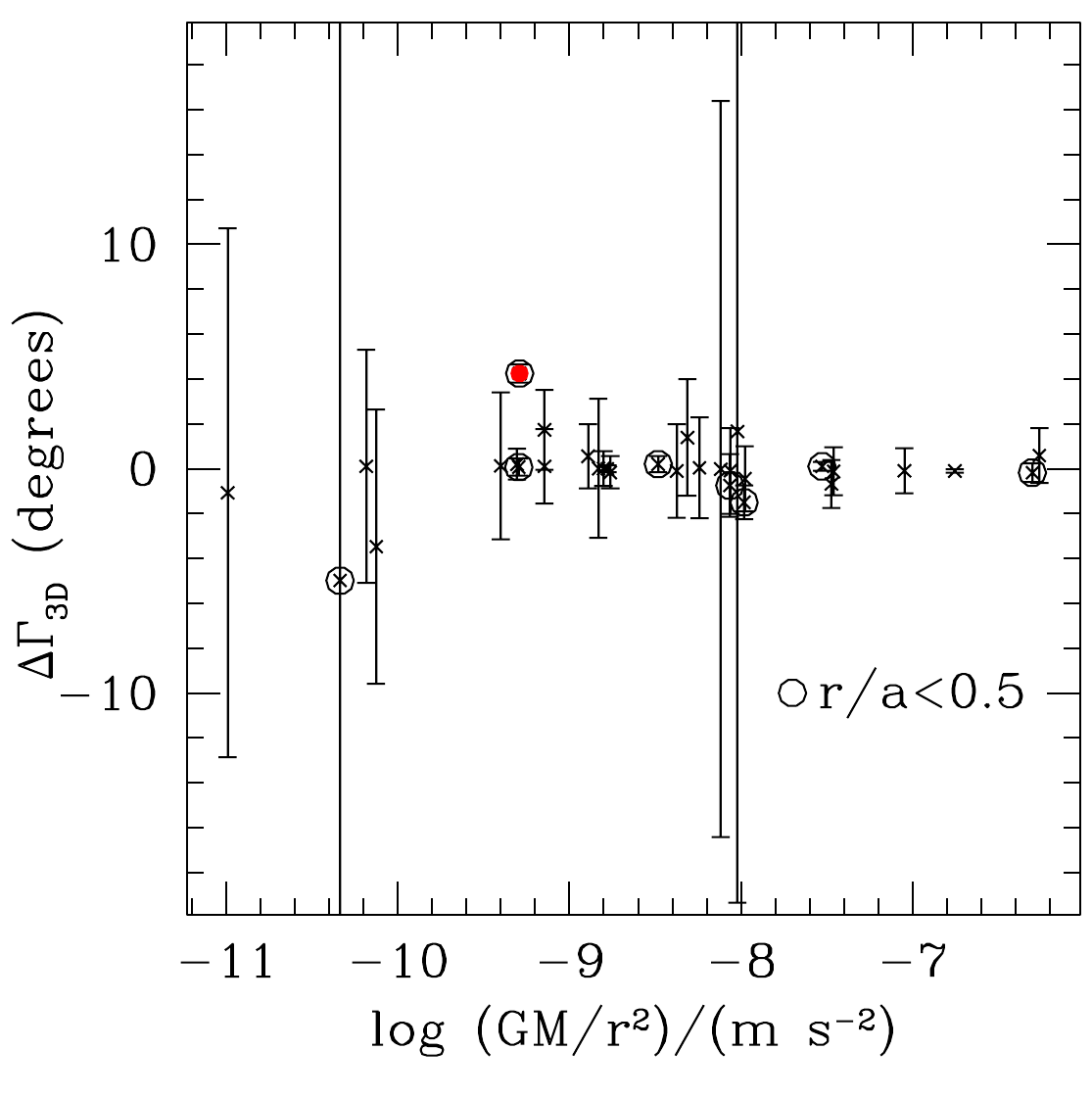}
  \caption{The residuals $\Delta V_{\rm rad}/V_{\rm tot}$ (left),
    $\Delta \Gamma_{\rm 2D}$ (middle), and $\Delta \Gamma_{\rm 3D}$ (right) as a function
    of acceleration. We circle WBs with inclinations larger than 70$^\circ$ in the left, WBs with eccentricities larger than 0.8 in the middle, and WBs with $r/a<0.5$ in the right plot. }
    \label{fig_fitquality}
\end{figure*}

\begin{figure*}
  \includegraphics[width=0.33\linewidth]{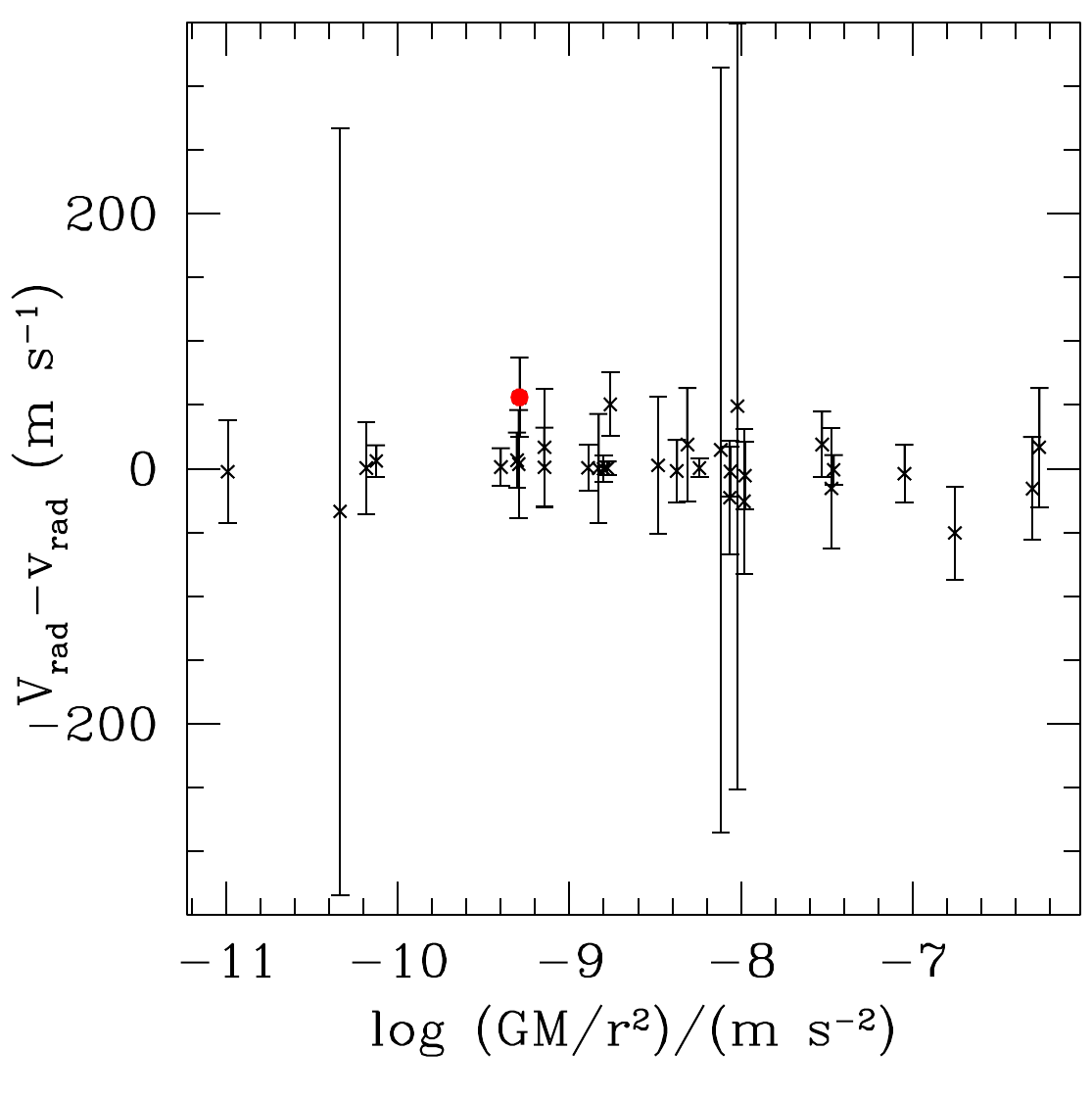}
  \includegraphics[width=0.33\linewidth]{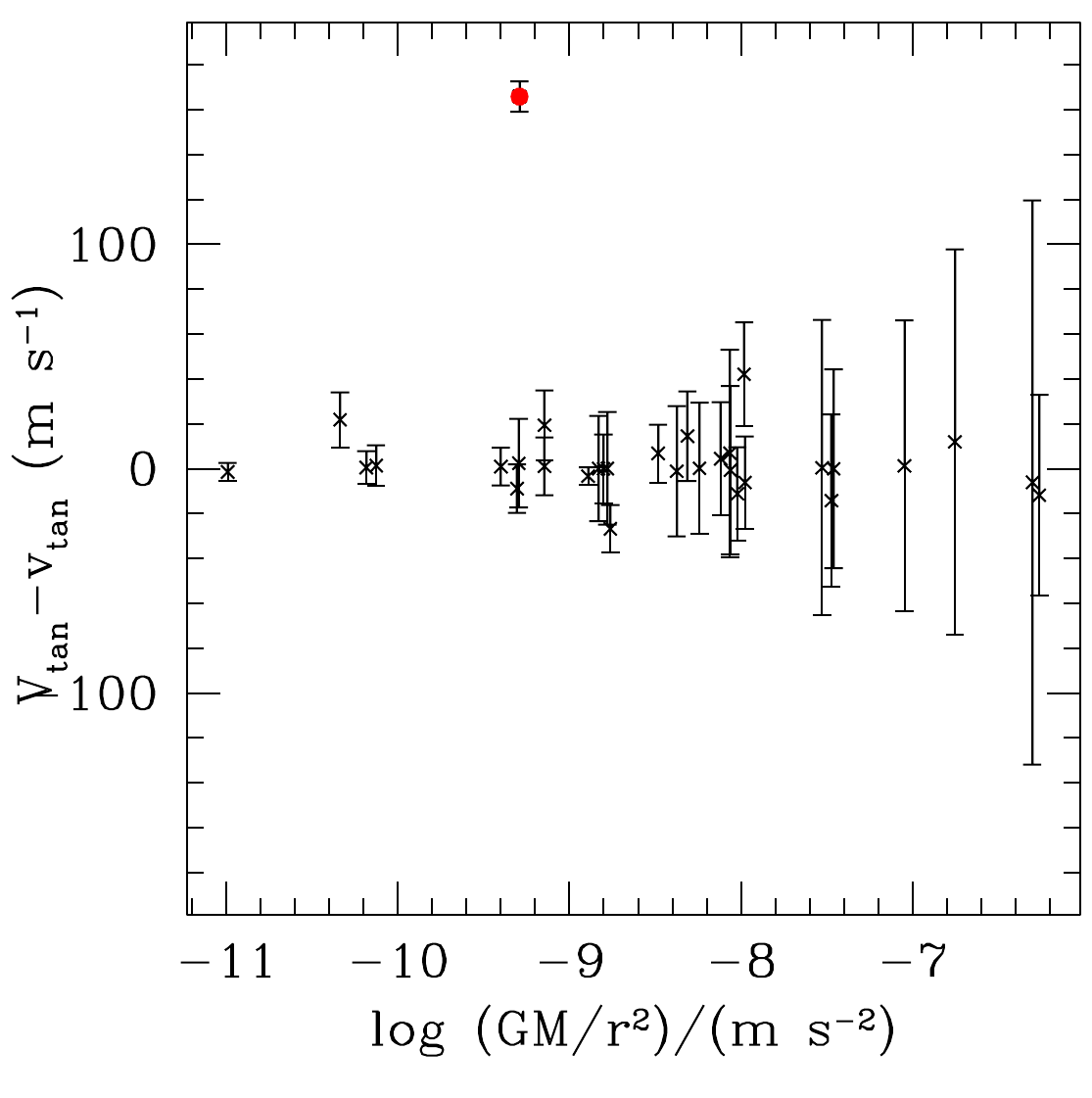}
    \includegraphics[width=0.33\linewidth]{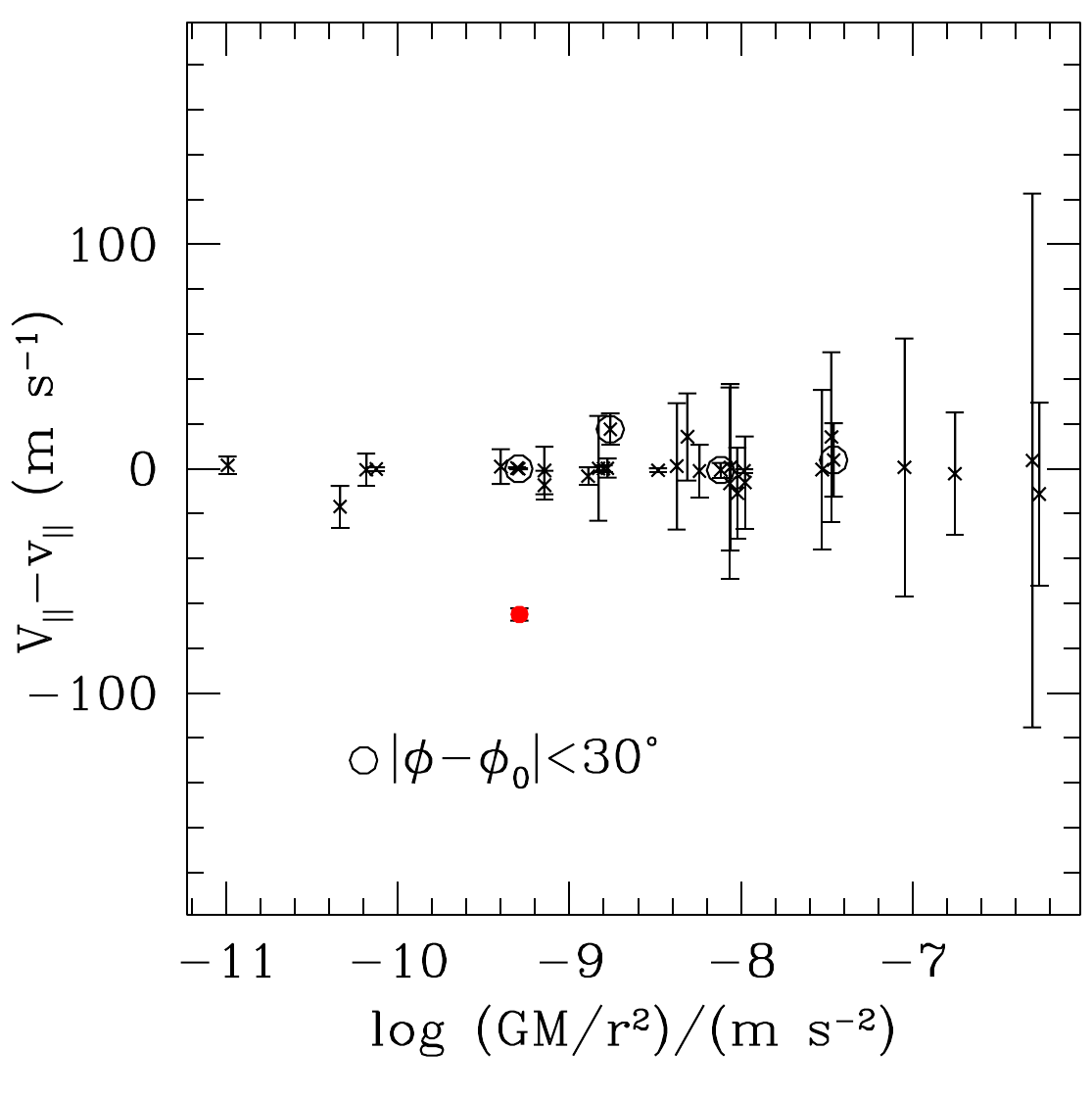}
  \caption{The residuals $V_{\rm rad}-v_{\rm rad}$ (left),
    $V_{\rm tan}-v_{\rm tan}$ (middle), and $V_\parallel-v_\parallel$ (right) as a function
    of acceleration.}
    \label{fig_fitVquality}
\end{figure*}

\begin{figure}
  \includegraphics[width=\linewidth]{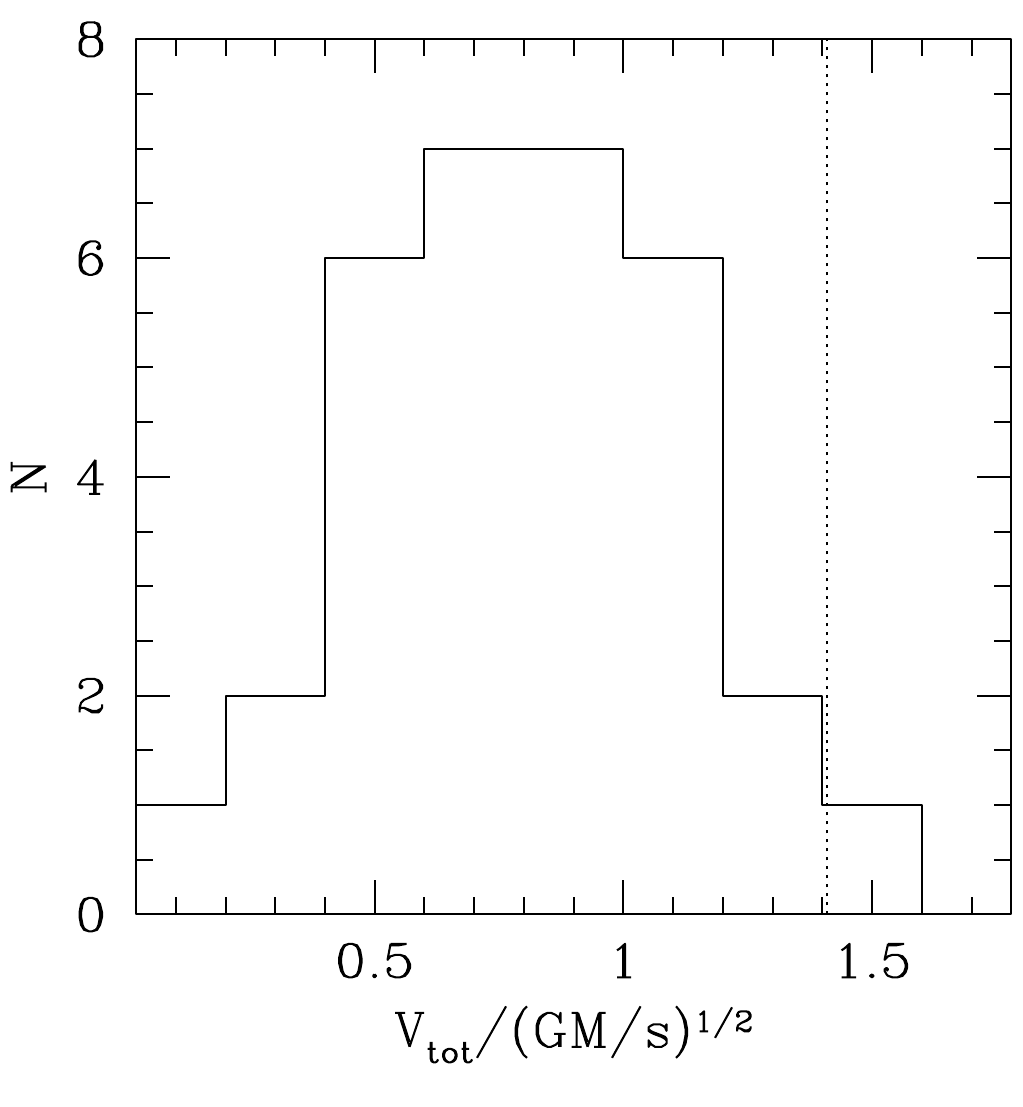}
  \caption{The histogram of the ratio $\tilde{V}=V_{\rm tot}/V_{\rm circ,s}$;
    except for WB \#24, $\tilde{V}$ is always less than $\sqrt{2}$ (dotted vertical line),
    as expected in Newtonian dynamics.}
    \label{fig_hisVtotVcirc}
\end{figure}

The next two figures assess the quality of our best fit solutions.
Fig. \ref{fig_fitquality} shows that the ratio $V_{\rm rad}/V_{\rm
  tot}$, the angle $\Gamma_{\rm 2D}$ and the angle $\Gamma_{\rm 3D}$
are well reproduced within the errors. The residuals do not show
trends with acceleration or separation $r$. Fig. \ref{fig_fitVquality}
presents the same information in terms of velocity residuals, where we
have added in quadrature an error equal to 3\% of the model values to
take into account the uncertainties on the total mass of the
binaries. Without WB \#24, radial, tangential and parallel velocities
are reproduced with RMS of 21, 12 and 7 m~s$^{-1}$,
respectively. Total velocities are reproduced exactly by
construction. This investigation demonstrates that Newtonian solutions
can be found down to the lowest accelerations probed by our (small)
sample of WBs. Furthermore, Fig. \ref{fig_hisVtotVcirc} shows the
histogram of the ratio $\tilde{V}=V_{\rm tot}/V_{\rm circ,s}$, where
$V_{\rm circ,s}=(GM/s)^{1/2}$ is the Newtonian circular velocity for
the current projected separation. As expected, all WBs except binary
\#24 have $\tilde{V}<\sqrt{2}$ and in qualitatively agreement with the
Newtonian distributions of \citet{Pittordis2018}.

In the following we discuss how plausible these (non-unique) solutions are.

The circled WBs in Fig. \ref{fig_fitquality} show that some of the large
separation, low-acceleration WBs have inclinations larger than
70$^\circ$, or eccentricities larger than 0.8, or separations smaller
than half the semi-major axis, or phases within 30$^\circ$ of the
pericenter. A larger sample of WBs at large separations is needed to assess
whether these possibly extreme properties persist.

\begin{figure}
  \includegraphics[width=\linewidth]{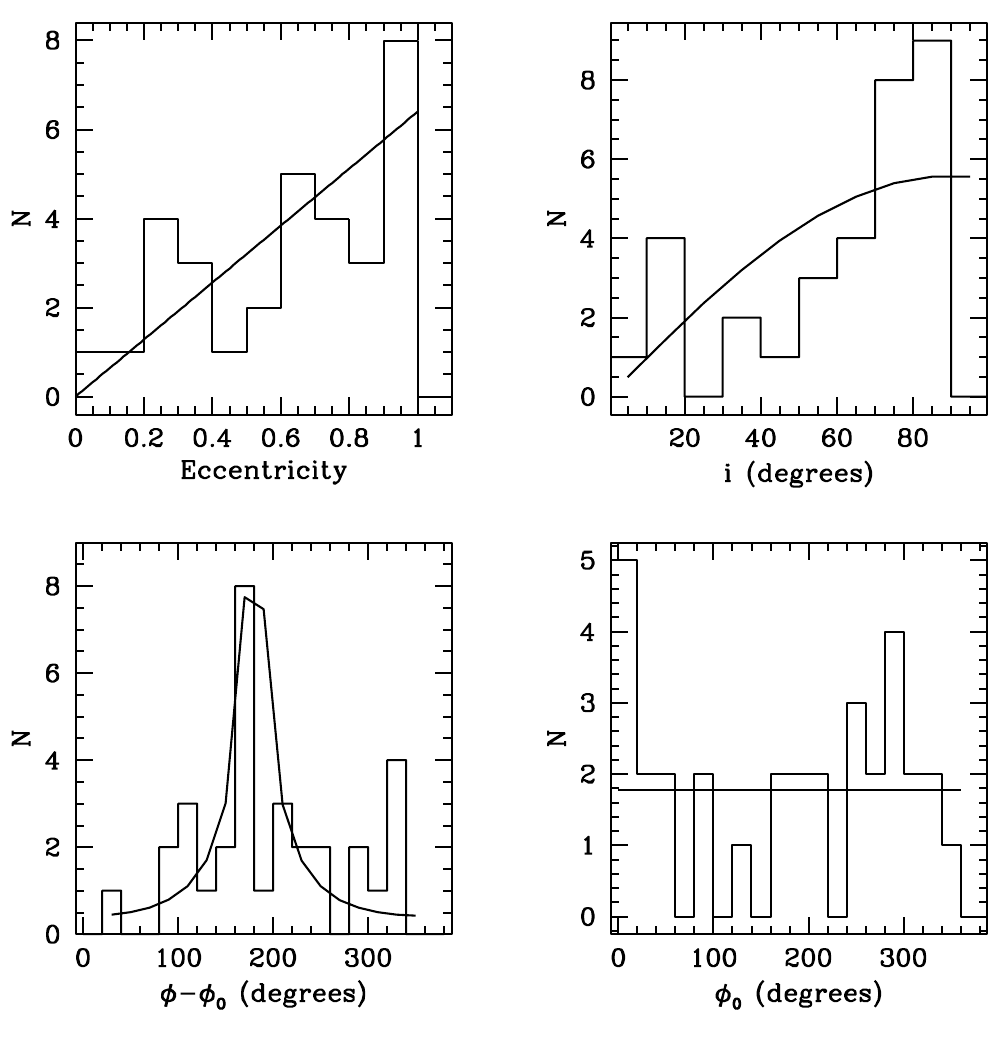}
  \caption{The distributions of best-fitting parameters. The histogram of best-fitting of eccentricities (top left), of inclinations (top right), of phases $\phi-\phi_0$ (bottom left) and of orientation angles $\phi_0$ (bottom right) with the expected distributions (black lines).
}
  \label{fig_hiseiphi0}
\end{figure}

Figure \ref{fig_hiseiphi0} shows the histograms of best-fitting
eccentricities, inclination, phase angles $\phi-\phi_0$, and semi-major
orientation angles $\phi_0$. The eccentricity distribution is
approximately thermal ($f(e)de=2e de$), as expected for WB with
separations larger than 1 kAU \citep{Tokovinin2020}. The inclination
distribution is peaked towards higher values, to some extent exceeding
the expectations from a random distribution of inclinations (flat in
$\cos i$). Instead of the expected 15 $\pm$ 4 WBs with inclinations
larger than 60$^\circ$, we count 21 $\pm 4$ such systems, possibly
still within the range allowed by low-number statistics. The phase
distribution has the expected peak at apocenter and a suspicious
second peak at pericenter exceeding expectations, but again with low
number statistics. We compute the distribution of expected phases
using Eq. \ref{eq_tT}, averaging over the best-fitting
eccentricities. The
orientation of the semi-major axis is approximately random, as
expected.

\begin{figure}
  \includegraphics[width=0.49\linewidth]{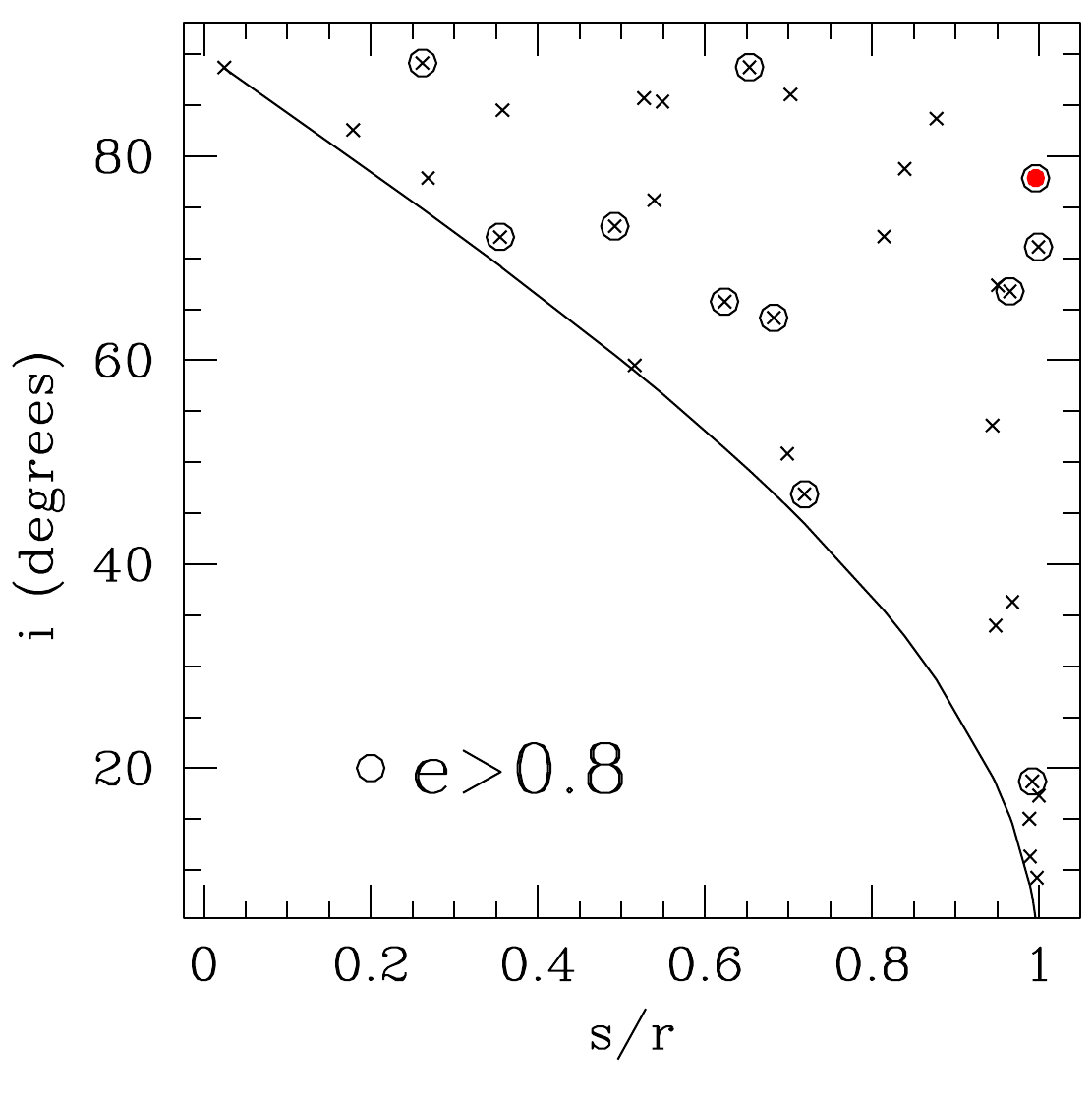}
  \includegraphics[width=0.49\linewidth]{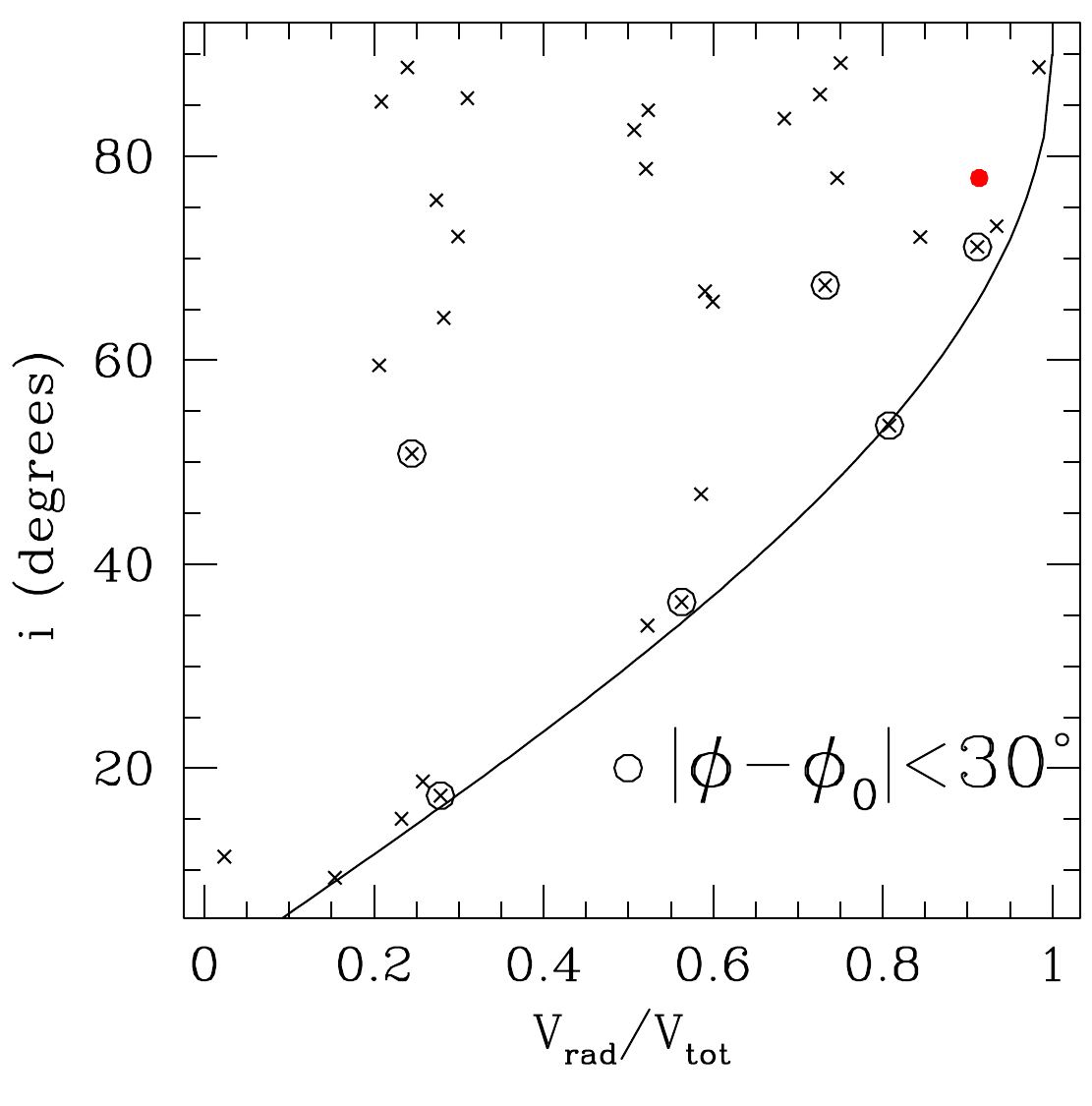}
  \includegraphics[width=0.49\linewidth]{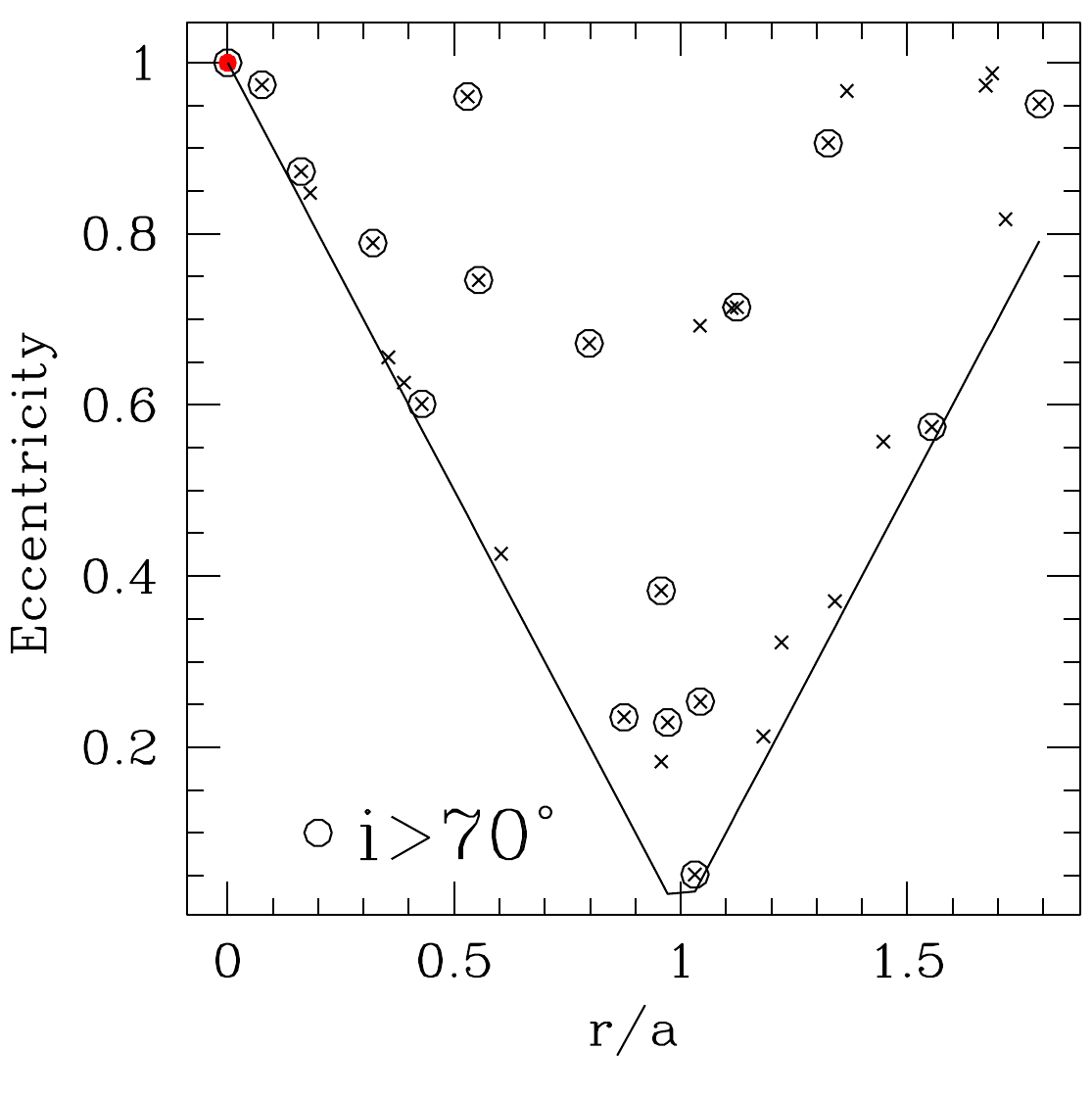}
  \includegraphics[width=0.49\linewidth]{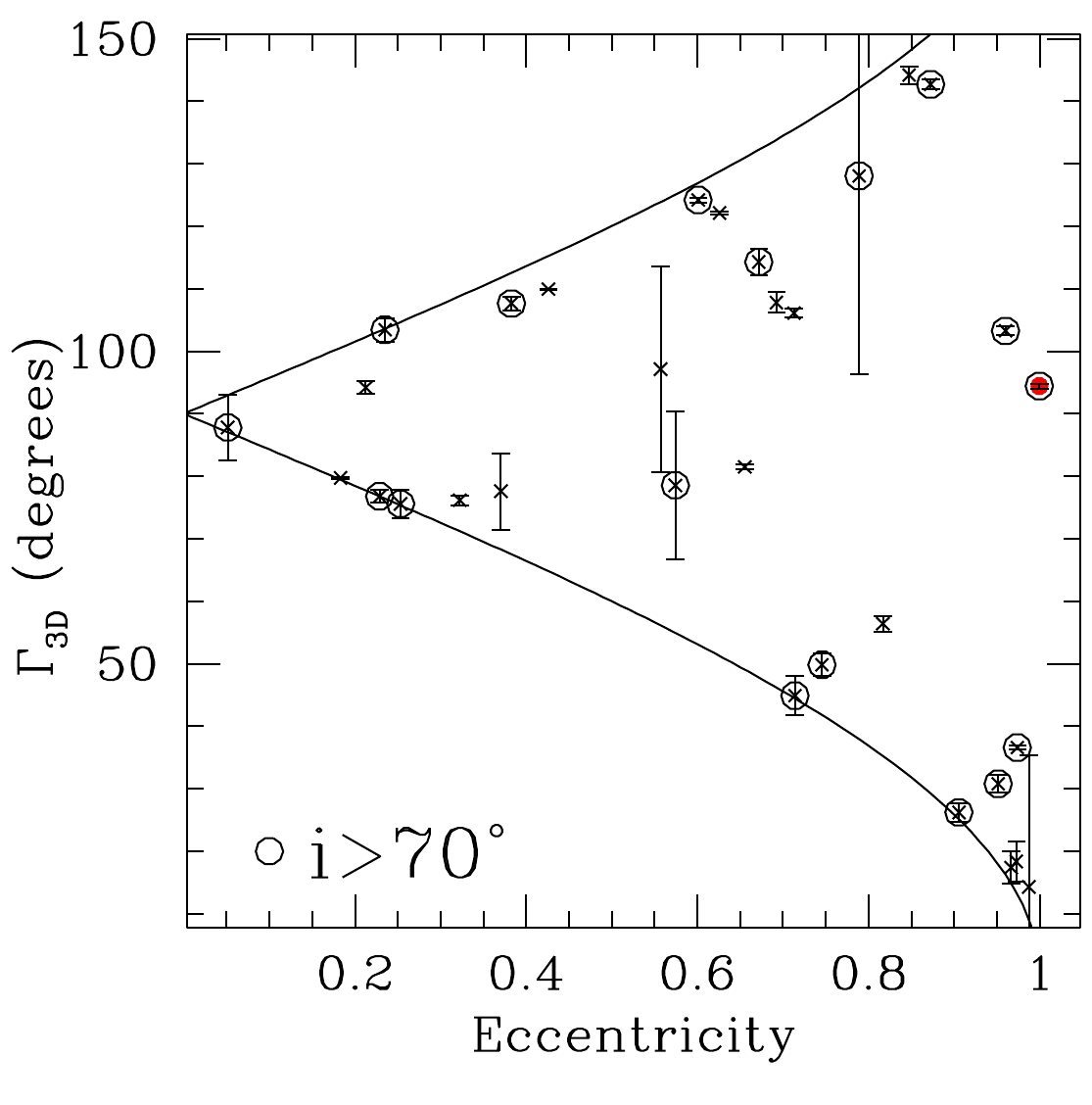}  
\caption{Top left: inclination as a function of $s/r$. The line shows
  the minimum allowed inclination. The circled WBs have eccentricities
  larger than 0.8.  Top right: inclination as a function of
  $V_{\rm rad}/V_{\rm tot}$.  The line shows the minimum allowed
  inclination. The circled WBs have phases near the pericenter within
  30$^\circ$. Bottom left: eccentricity as function of $r/a$. The line
  shows the minimum allowed eccentricity. The circled WBs have
  inclinations larger than 70$^\circ$. Bottom right: $\Gamma_{\rm 3D}$ as
  a function of eccentricity. The lines show the allowed angle
  range. The circled WBs have inclinations larger than 70$^\circ$.}
  \label{fig_IncEcc}
\end{figure}

Figure \ref{fig_IncEcc} shows how the ratios $s/r$, $V_{\rm rad}/V_{\rm tot}$
and $r/a$ constrain the inclination and eccentricity of the
binaries. Two of the high eccentricity WBs are forced to have high
inclinations ($\ge 70^\circ$) by the required low values of $s/r$, further two by the
large $V_{\rm rad}/V_{\rm tot}$ ratio. WBs requiring low or high values of
$r/a$ are forced to have high eccentricities. Five of them have high
inclinations, of which one is the possibly unbound pair \#24.
The measured angle $\Gamma_{\rm 3D}$ forces eight WBs to have
large eccentricities
($\ge 0.8$),
of which four have high inclination.
Four of the WBs with phases near pericenter within 30$^\circ$ have the
minimum inclination allowed by the $V_{\rm rad}/V_{\rm tot}$ ratio.

\begin{figure}
  \includegraphics[width=\linewidth]{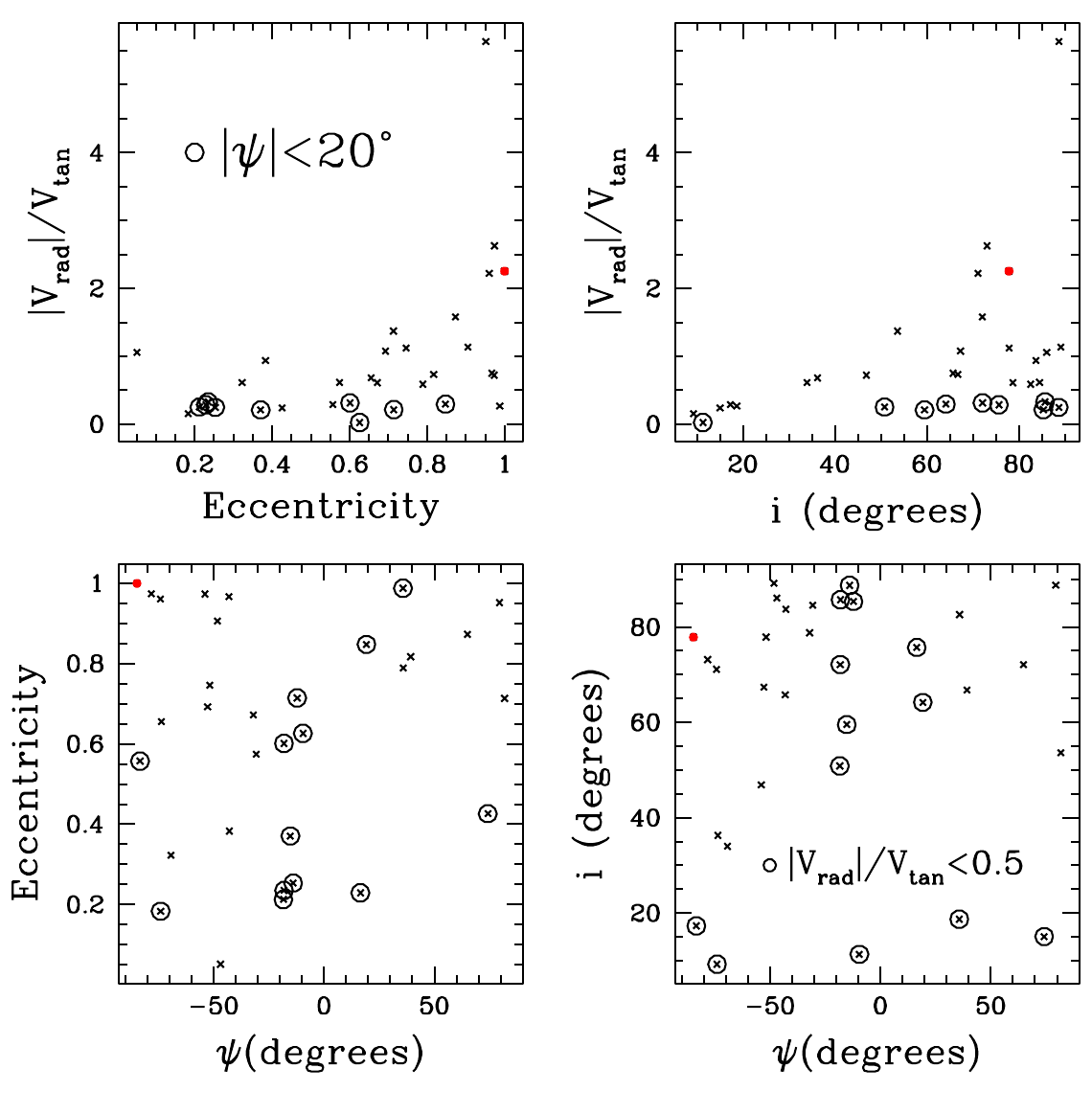}
 \caption{Correlations with $|V_{\rm rad}|/V_{\rm tan}$ (top) and $|\psi|$ (bottom): eccentricity (left panels) and inclination (right panels).  Circled crosses in the top panels show binaries with $|\psi|<20^\circ$; circled crosses in the bottom panels show binaries with $|V_{\rm rad}|/V_{\rm tan}<0.5$.}
  \label{fig_VradVtan}
\end{figure}

Figure \ref{fig_VradVtan} investigates the role of the angle $\psi$ to
reproduce the ratio between the radial and tangential velocities
(see Eq. \ref{eq_vradvtan}).
The top panels show, as expected, that the ratio
$|V_{\rm rad}|/V_{\rm tan}$ is small for small eccentricities and inclinations
and on average larger for large inclinations, with the largest values
having also large eccentricities. The bottom panels show that low values of
$|V_{\rm rad}|/V_{\rm tan}$ at
high eccentricities and inclinations are due to low values of $|\psi|$.

\begin{figure}
  \includegraphics[width=\linewidth]{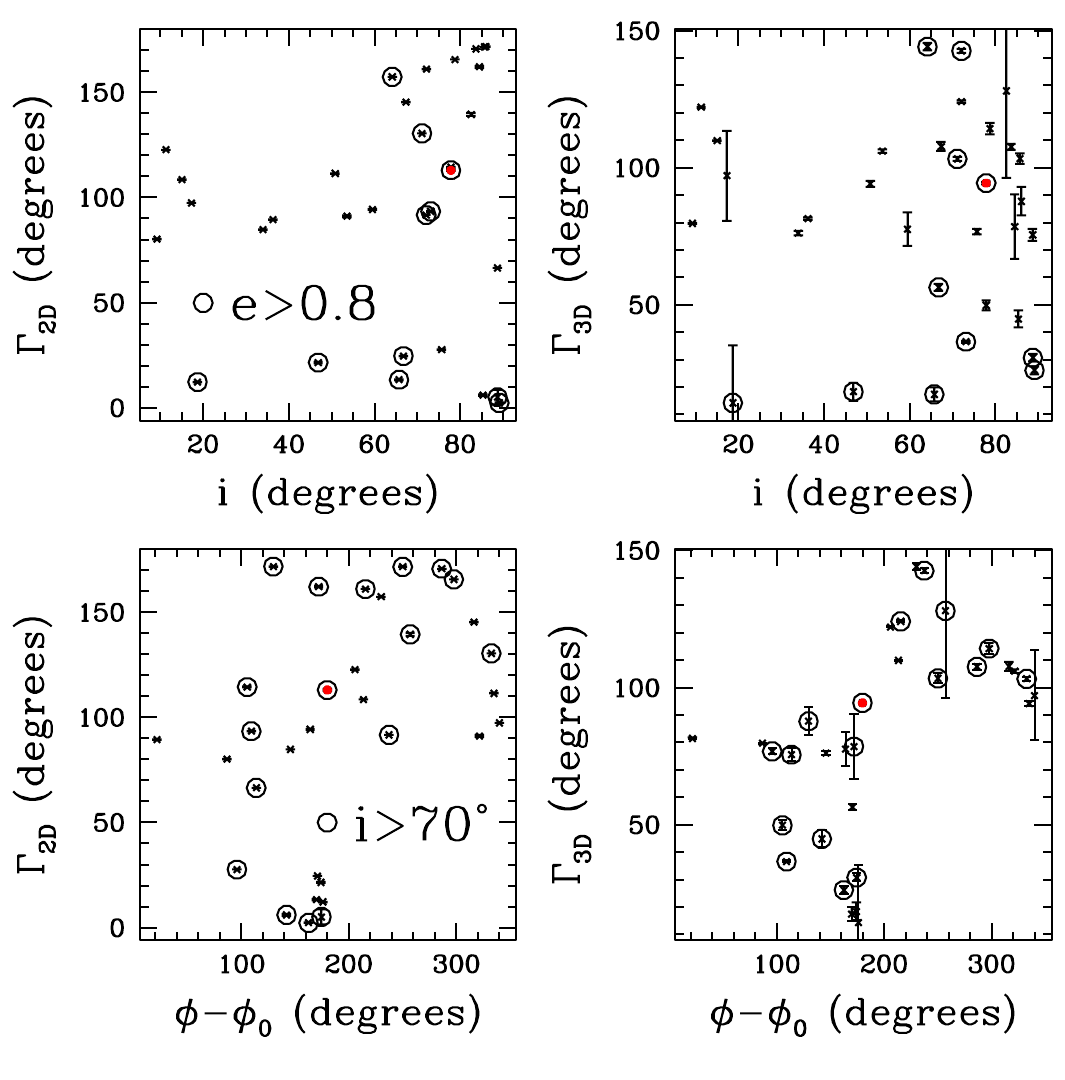}
  \caption{Correlations with $\Gamma_{\rm 2D}$ (left) and $\Gamma_{\rm 3D}$ (right).
    The top plots show both angles as a function of inclination, circled
  WBs have eccentricities larger than 0.8. The bottom plots show both angles as function of orbit phase, circled WBs have inclinations larger than 70$^\circ$.}
  \label{fig_Gamma2D3D}
\end{figure}

Figure \ref{fig_Gamma2D3D} show correlations with $\Gamma_{\rm 2D}$ and
$\Gamma_{\rm 3D}$.  As expected, both angles are $\approx 90$ degrees at
lowish inclinations and eccentricities. Consistently, the sky position
and velocity vectors tend to be parallel or anti-parallel at high
inclinations and high eccentricities. While no obvious trend between
$\Gamma_{\rm 2D}$ with $\phi-\phi_0$ appears (bottom left of Fig.
\ref{fig_Gamma2D3D}), the expected modulation (see
Eq. \ref{eq_gamma3d}) of $\Gamma_{\rm 3D}$ with $\phi-\phi_0$, scaled by
the eccentricity value is clearly seen in the bottom
right plot.

Finally, we summarize further (absence of) trends with eccentricities,
inclinations, orbital phases, semi-major position angles and
separations. There is a lack of eccentricities larger than 0.6 with
inclinations lower than 45 degrees. There is a lack of low
eccentricities at large separations, and a lack of high eccentricities
at smaller separations, in line with the tendency discussed by
\citet{ElBadry2021} of WBs having larger $e$ at larger
separations. There are no obvious correlations between eccentricity or
inclination and phase $\phi-\phi_0$ or semi-major axis position
$\phi$.

\section{Conclusions} 
     \label{sec_conclusions}

In this work we have introduced two main novelties: the use of
accurate radial velocities and the exploitation of the distance
distribution provided by the Gaia parallaxes.  With these, we are able
to constrain the orbits of the WBs to test Einstein-Newton gravity in
the low potential regime.  We searched for archival data of WB
candidates observed with the high precision spectrograph HARPS,
finding 44 systems.

Precise RVs, repeated in time, allow to
single out and remove multiple systems, and high resolution observations 
deliver precise chemical abundances and stellar
parameters. After cleaning, 32 out of the initial 44 systems  (73$\%$) survive
our conservative selection criteria and these systems are
suitable for testing Einstein-Newton law.  The final sample has a
median RUWE of 0.975 (with a highest value of 1.24) and a RV
median variability of 10$\pm$6 ms$^{-1}$. The star with the highest RV
variability (29 m~s$^{-1}$), is known to host a massive exoplanet.

We obtain for each star an accurate RV, correcting the
precise HARPS shifts for gravitational redshift and for convective
motions in the stellar atmospheres using 3D model atmospheres.  The
corrected velocities are accurate to better than 40 m~s$^{-1}$.  The
comparison with Gaia shows a remarkable agreement in the RV zero point
(67 m~s$^{-1}$) with a spread of 311 m~s$^{-1}$ .

We exploit the distance distributions provided by the Gaia parallaxes
to probe projected and tri-dimensional separations and velocity
differences.  In 31 of the 32 WBs at least one Newtonian orbit
solution is possible with distances compatible with the Gaia parallax
errors, the precision of the total mass estimates and the small
velocity errors. For these 31 WBs we obtain reasonable orbit
parameters, with some systems possibly too near pericenter and/or at
too high inclination. We
suspect that the remaining WB, \#24, is not bound. 

\citet{Chae+2024, Chae2025} finds a remarkable difference between WBs
with separations larger than $\sim$2 kAU (0.01 pc), and the closer
systems, with the widely separated WBs breaking Newton law. We do not
detect such a trend with our WB sample, much smaller in size but with
higher precision radial velocities. To check whether this conclusion
is statistically robust, we are in the process of collecting similarly
good RVs for a large sample of WBs, to not only confirm the existence
of Newtonian solutions in all cases, but also to verify the
distributions of the retrieved orbital parameters are physically
acceptable.

\begin{acknowledgements}
We thank the referee Charalambos Pittordis for a constructive report that helped
us improving the presentation of our results.  MTM acknowledges the
support of the Australian Research Council through Future Fellowship
grant FT180100194. Research activities of the Board of Observational
and Instrumental Astronomy (NAOS), at the Federal University of Rio
Grande do Norte, are supported by continuous grants from the Brazilian
funding agency Conselho Nacional de Desenvolvimento Científico
(CNPq). I.C.L. (grant No. 313103/ 2022-4), and J.R.M. (grant
No. 308928/2019-9) acknowledge CNPq research fellowships. This work
has made use of data from the European Space Agency (ESA) mission {\it
  Gaia} (\url{https://www.cosmos.esa.int/gaia}), processed by the {\it
  Gaia} Data Processing and Analysis Consortium (DPAC,
\url{https://www.cosmos.esa.int/web/gaia/dpac/consortium}).  Funding
for the DPAC has been provided by national institutions, in particular
the institutions participating in the {\it Gaia} Multilateral
Agreement.
\end{acknowledgements}

\bibliographystyle{aa} 
\bibliography{wb4.bib} 

\begin{appendix}
\section{Systems discarded from original sample}
\label{App_discarded}

Out of the 44 initial systems with HARPS RVs, 12 have been discarded as
unsuitable for the present study. We provide the interested reader
with the justification of our choice.

\begin{description}
\item[HD20430 and HD20439:] Our analysis provides an age of $\sim$700
Myrs. These stars belong to the Hyades cluster and may just be a proper
motion pair.

\item[HD13904 and BD10303B:] the last observations in the catalogue of both
stars give a difference of -0.6 km~s$^{-1}$ with respect to the
previous observations. The observations have been re-reduced and the
change in radial velocity confirmed. This seems a quadruple
system. HD13904 has a RUWE larger than 2.

\item[HD11790:] RV varies of 500 m~s$^{-1}$ in a few days, the star has a
companion.

\item[CD-261686 and CD-2616866B:] there is some confusion amongst
  the two target in the
HARPS observations, and a difference between them of 3.7 km~s$^{-1}$
in radial velocity, that is in sharp disagreement with the RV
difference measured by Gaia for these stars. Such a large RV
difference is not compatible with a bound binary and the disagreement
with Gaia suggests that the system is multiple.

\item[HIP34407 and HD54100:] \citet{Spina+2021} report a difference of 0.15 dex
in the [Fe/H] of the two stars. \citet{Ramirez2019} notice a spread of 60
m~s$^{-1}$ in radial velocity in their observations, indicating a
possible companion.  HARPS spectra (5 and 2 observations respectively)
do not show high RV spread.

\item[HD18142 and HD181544:] HD 181544 shows a 30 m~s$^{-1}$ variability on 7
exposures, that is higher than average, and has a RUWE of 1.64.

\item[HD2567 and HD2638:] many good HARPS observations are available. HD2638
has a high RV
variability (47 m~s$^{-1}$) but is known to host a Jupiter-planet 
\citep{Moutou2005}.  The system is however triple, with an M star
close to HD2638 \citep{Wittrock2016}, so not suitable for the test.

\item[HD169392A and B:] HD169392A has a RUWE of 4.2. 

\item[HD182817 and HD182797:] HD182817 has a RUWE of 1.6 and a RV variability
of 35 m~s$^{-1}$ .

\item[HD116920 and HD116858:] HD116920 has a RV variability of 111 m~s$^{-1}$.
\end{description}

\onecolumn  

\begin{table*}
\section{Tables}
\label{app_Table}
\caption{Gaia parameters of the 64 WB stars.} 
\label{tab_Gaia} 
\centering
{\small
\begin{tabular}{r r r r r r r r}
  \hline
  \hline
  Nr. & Star & $\alpha$~~~~~~ & $\delta$~~~~~~ & Parallax~~~~~ & pm$\alpha$~~~~~~~~~ & pm$\delta$~~~~~~~~~~ & RUWE \\
 & & (degrees) & (degrees) & (mas)~~~~~~~ & (mas~yr$^{-1}$)~~~~~ & (mas~yr$^{-1}$)~~~~~~ & \\
  \hline
 1 & HD4552 &   11.92189 &   12.87912 &      9.188 $\pm$      0.016 &     84.485 $\pm$      0.021 &    -35.236 $\pm$      0.016 & 0.81 \\    
 1 & BD+120090 &   11.93224 &   12.87200 &      9.229 $\pm$      0.017 &     84.098 $\pm$      0.020 &    -35.772 $\pm$      0.015 & 0.90 \\ 
 2 & HIP6130 &   19.67317 &   -0.86869 &     36.399 $\pm$      0.025 &    432.564 $\pm$      0.032 &   -253.357 $\pm$      0.026 & 0.91 \\   
 2 & G56.3B &   19.66951 &   -0.87553 &     36.343 $\pm$      0.031 &    435.477 $\pm$      0.043 &   -254.683 $\pm$      0.031 & 0.88 \\    
 3 & CD-50524 &   28.10670 &  -49.53443 &      8.598 $\pm$      0.013 &     32.434 $\pm$      0.012 &    -48.125 $\pm$      0.013 & 0.99 \\  
 3 & HD11584 &   28.11510 &  -49.52397 &      8.605 $\pm$      0.013 &     31.377 $\pm$      0.012 &    -47.560 $\pm$      0.013 & 1.01 \\
\hline

\end{tabular}
}
\end{table*}

\begin{table*}
  \caption{Stellar parameters and radial velocities of the 64 WB stars.
  The gravity values given in italics are Gaia values.} 
\label{tab_Stars} 
\centering
{\tiny
\begin{tabular}{r r r r r r r r r r r r}
  \hline
  \hline
  Nr. & Star & Age & T$_{\rm eff}$~~~~ & log~g~~~~~~~ & [Fe/H]~~~~~~ & Mass & Radius & GR~~ &  $V_{\rm con}$ & Mask/N & RV~~~~~~~~~~~\\
      &      & (Gyr) & (K)~~~~~ & (m~s$^{-2}$)~~~~~   & (dex)~~~~~~~  & $(M_\odot)$ & $(R_\odot)$ & (km~s$^{-1}$) & (km~s$^{-1}$) & & (km~s$^{-1}$)~~~~~~~~\\
  \hline
 1                   &      HD4552           &         2.6           &        6278       $\pm$          23           &       4.323       $\pm$       0.056           &      -0.005       $\pm$       0.016           &        1.15           &        1.13           &        0.65           &       -0.67           &       G2/13           &      14.761       $\pm$       0.005          \\  
 1                   &   BD+120090           &         6.3           &        5994       $\pm$           6           &       4.298       $\pm$       0.016           &      -0.011       $\pm$       0.005           &        1.03           &        1.18           &        0.56           &       -0.59           &       G2/13           &      15.219       $\pm$       0.016          \\  
 2                   &     HIP6130           &        13.8           &        5300       $\pm$          50           &        4.43       $\pm$         0.1           &       -0.13       $\pm$        0.05           &         0.8           &        0.79           &        0.64           &       -0.39           &       K5/11           &      17.140       $\pm$        0.01          \\  
 2                   &      G56.3B           &         5.1           &        3930       $\pm$         100           &        {\it 4.40}           &        9999       $\pm$        9999           &        0.55           &        0.49           &        0.71           &       -0.32           &        K5/2           &      18.416       $\pm$       0.002          \\
\hline

\end{tabular}
}
\end{table*}

\begin{table*}
\caption{Angular separations, masses and recovered distances of the 32 WBs.} 
\label{tab_chi} 
\centering                         
\begin{tabular}{cllrrrrrrr}
  \hline
  \hline
  Nr. & Star 1 & Star 2 & $(\alpha_1-\alpha_2)\cos \overline{\delta}$ & $\delta_1-\delta_2$ &  $\theta$~~~~ & M~~~  &$D_1$~~~  & $D_2$~~~ & $\chi^2_{\rm D}$ \\
   & & & $('')$~~  & $('')$~~ & $('')$~~ & ($M_\odot$) &  (pc)~~ &  (pc)~~ & \\
\hline
   1                &      HD4552           &   BD+120090           &       36.31           &      -25.65           &       44.46           &        2.18           &     108.591           &     108.583           &       3.033          \\  
   2                &     HIP6130           &      G56.3B           &       13.17           &       24.60           &       27.90           &        1.35           &      27.490           &      27.497           &       1.435          \\  
   3                &    CD-50524           &     HD11584           &       19.62           &       37.64           &       42.45           &        2.22           &     116.265           &     116.271           &       0.154          \\  
   4                &    HD13357A           &    HD13357B           &        5.38           &        2.69           &        6.02           &        1.73           &      42.912           &      42.917           &       0.008          \\  
   5                &     HD20766           &      HR1010           &      184.23           &      248.21           &      309.11           &        1.85           &      12.045           &      12.039           &       0.004          \\  
   6                &     HD20781           &     HD20782           &        8.35           &     -252.85           &      252.99           &        1.84           &      35.928           &      35.884           &       0.991          \\
\hline
\end{tabular}
\end{table*}

\begin{table*}
\caption{Measured velocities and angles of the 32 WBs.} 
\label{tab_binaries} 
\centering                         
\begin{tabular}{crrrrrrr}
  \hline
  \hline
Nr. & $V_{\rm ra}$~~~~~~~ & $V_{\rm dec}$~~~~~~~ & $V_{\rm rad}$~~~~~~~ & $V_{\rm tan}$~~~~~~~ & $V_{\rm tot}~~~~~~~$ & $\Gamma_{\rm 2D}$~~~~~~~~  & $\Gamma_{\rm 3D}$~~~~~~~~\\
&  (m~s$^{-1}$)~~~~~ & (m~s$^{-1}$)~~~~~ & (m~s$^{-1}$)~~~~~ & (m~s$^{-1}$)~~~~~ & (m~s$^{-1}$)~~~~~ & (degree)~~~~ & (degree)~~~~  \\
\hline
  1                  &      -202.4       $\pm$         3.1           &      -274.8       $\pm$         2.4           &       468.0       $\pm$        16.8           &       341.3       $\pm$         2.8           &       579.2       $\pm$        13.3           &        91.1       $\pm$         0.5           &       106.1       $\pm$         0.7          \\  
  2                  &      -393.2       $\pm$         1.5           &       180.7       $\pm$         1.1           &     -1136.0       $\pm$        41.3           &       432.7       $\pm$         1.5           &      1215.6       $\pm$        34.8           &        93.5       $\pm$         0.1           &        36.6       $\pm$         0.4          \\  
  3                  &       581.6       $\pm$         2.0           &      -309.6       $\pm$         2.1           &      -449.0       $\pm$         40.2           &       658.9       $\pm$         2.1           &       797.3       $\pm$         23.0           &        89.5       $\pm$         0.2           &        81.5       $\pm$         0.4        \\  
  4                  &       617.6       $\pm$         1.7           &       277.2       $\pm$         1.6           &      -769.0       $\pm$        13.4           &       677.0       $\pm$         1.7           &      1024.5       $\pm$        10.4           &         2.4       $\pm$         0.1           &        26.2       $\pm$         1.5          \\
\hline
\end{tabular}
\end{table*}

\begin{table*}
\caption{Orbit-based parameters of the 32 WBs.} 
\label{tab_modelBest} 
\centering                         
{\tiny
\begin{tabular}{rrrrrrrrrrrrrrrr}
  \hline
  \hline
Nr. & $s$~~~~ & $r$~~~~ & $a$~~~~ & ~~$e_{\rm min}$ & $e$~~ & $i_{\rm min}$~ & $i$~~~~ & $\phi-\phi_0$ &
  $\phi_0$~~~~ & $\psi$~~~~ & $\gamma_{\rm 2D}$~~ & $\gamma_{3\rm D}$~~ & log P &
  log $A_{\rm N}$\\
    &  (pc)~~ & (pc)~~ & (pc)~~ & & & (degree) & (degree) & (degree) & (degree) & (degree) & (degree) & (degree) & (yr) & (m~s$^{-2}$) \\
\hline
  1                  &      0.0234           &      0.0248           &       0.022           &        0.11           &        0.71           &        19.2           &        53.6           &       321.5           &        14.3           &        81.7           &        91.2           &       105.9           &        5.32           &       -9.31          \\   
  2                  &      0.0037           &      0.0076           &       0.099           &        0.92           &        0.97           &        60.5           &        73.1           &       109.4           &       136.1           &       -78.2           &        93.4           &        36.4           &        6.40           &       -8.48          \\   
  3                  &      0.0239           &      0.0247           &       0.070           &        0.65           &        0.66           &        14.5           &        36.3           &        21.8           &       183.3           &       -73.5           &        89.5           &        81.4           &        6.04           &       -9.25          \\   
  4                  &      0.0013           &      0.0048           &       0.004           &        0.33           &        0.91           &        74.8           &        89.1           &       162.6           &       302.5           &       -48.2           &         2.2           &        26.7           &        4.19           &       -7.98          \\
\hline

\end{tabular}
\tablefoottext{a}{Mass increased by 33\%.}
}
\end{table*}

\end{appendix}

\end{document}